\documentclass[final,3p,times,number,sort&compress]{elsarticle}
\usepackage{amssymb}
\usepackage{amsthm}
\usepackage{amsmath}
\usepackage{url}

\journal{Physica A}

\begin{document}

\begin{frontmatter}



\title{Theoretical conditions for restricting secondary jams in jam-absorption driving scenarios}


\author[label1]{Ryosuke Nishi\corref{cor1}}
\ead{nishi@tottori-u.ac.jp}
\cortext[cor1]{Corresponding author. Tel.: +81 857 31 5192; fax: +81 857 31 5210.}
\address[label1]{Department of Mechanical and Aerospace Engineering, Graduate School of Engineering, Tottori University, 4-101 Koyama-cho Minami, Tottori 680-8552, Japan}

%arXiv: abstracts longer than 1920 characters will not be accepted
\begin{abstract}
There has been considerable interest in the active maneuvers made by a small number of vehicles to improve macroscopic traffic flows. Jam-absorption driving (JAD) is a single vehicle's maneuvers to remove a wide moving jam and consists of two actions. First, a vehicle upstream of the jam slows down and maintains a low velocity. Because it cuts off the supply of vehicles to the jam, the jam shrinks and finally disappears. Second, it returns to following the vehicle ahead of it. One of the critical problems of JAD is the occurrence of secondary jams. The perturbations caused by JAD actions may grow into secondary jams due to the instability of traffic flows. The occurrence of secondary jams was investigated by numerical simulations in non-periodic systems where only human-driven vehicles are placed upstream of the vehicle performing JAD. However, no theoretical condition has been proposed to restrict secondary jams in these systems. This paper presents a theoretical condition restricting secondary jams in a semi-infinite system composed of a vehicle performing JAD and the other human-driven vehicles obeying a car-following model on a non-periodic and single-lane road. In constructing this condition, we apply the linear string stability to a macroscopic spatiotemporal structure of JAD. Numerical simulations show that a finite version of this condition restricts secondary jams. Moreover, under this condition, we demonstrate that it is possible to restrict secondary jams in the semi-infinite system under wide ranges of the parameters of the system. Furthermore, we construct the conditions suppressing secondary jams in other semi-infinite systems with inflows from other lanes or a bottleneck, and demonstrate that JAD can restrict secondary jams in these systems. Thus, our method theoretically guarantees that a single vehicle can improve macroscopic traffic flows.
\end{abstract}


\end{frontmatter}


\section{\label{sec:introduction}Introduction}
%%%%%%%%%%%%%%%%%%%%%%%%%%%%%%%%%%%%%%%%%%%%%%%%%%%%%%%%%%%%%%%%%%%%%
As a collective phenomenon of self-driven particles, traffic jam has atrracted much attention of physists and its mechanism has been clarified dilligently~\citep{Chowdhury2000,Helbing2001,Kerner2004Physics,Schadschneider2010Stochastic}.
Traffic jam is also a huge social problem causing significant losses. For instance, automobile traffic jam caused the loss of 6.9 billion hours, 12 billion liters of fuel, and 160 billion U.S. dollars across 471 urban areas of the United States in 2014~\citep{Schrank2015}. Accordingly, there is a strong need to ease traffic jam.
For the case of highway traffic, many technologies have been developed for controlling traffic flow, which are categorized into two types. The first technologies do not need the devices mounted on vehicles but need the devices mounted on infrastructures: variable speed limits (VSL) using variable message signs~\citep{Lu2014, Khondaker2015}, and ramp metering (RM) using ramp meter signals~\citep{Papageorgiou2002}.
The second technologies need devices mounted on vehicles: adaptive cruise control (ACC)~\citep{Vahidi2003}, cooperative adaptive cruise control (CACC)~\citep{Arem2006,Dey2016}, and connected and/or automated vehicles~\citep{Li2018,Vahidi2018}.

Strategies to mitigate traffic jams on highways have also been developed diligently.
An efficient strategy is achieving a sufficient penetration ratio of connected and/or automated vehicles that contribute to the stabilization of traffic flows. For instance, ACC with its penetration ratio of $20-25\,\%$ can dissipate traffic jams through improving car-following performances of the vehicles equipped with it~\citep{Treiber2002,Davis2004,Kesting2008}.
More recent studies also support the improvement of traffic flows by connected and/or automated vehicles with a sufficient penetration ratio~\citep{Knorr2012,Wang2016IEEE,Gueriau2016,Talebpour2016}.
Other efficient strategy is controlling dynamically the traffic flow upstream and/or downstream of traffic jams~\citep{Hegyi2008,Carlson2010a,Nishi2013,VandeWeg2014,Jerath2015,Han2017,Stern2018}. This strategy is based on the fact that a traffic jam shrinks by restricting and enhancing the flow rate upstream and downstream of the jam, respectively.
This strategy uses the first technologies (such as VSL and RM)~\citep{Hegyi2008,Carlson2010a}, the second technologies (such as connected vehicles (CVs))~\citep{Jerath2015}, and the combinations of both technologies~\citep{VandeWeg2014,Han2017}.
Note that combinations of the first and the second strategies have also been investigated. For instance, tuning ACC parameters dynamically according to traffic situations mitigates traffic jams~\citep{Kesting2008}. Tuning microscopic parameters of a single connected and automated vehicle (CAV) according to traffic states also stabilizes a platoon of vehicles~\citep{Wang2018b}.
This paper focuses on the second strategy.

Controlling the traffic flow in the second strategy is realized by an appropriate manipulation of the spatiotemporal maneuvers of vehicles (such as changing their velocities at a specific time and position). This manipulation is categorized into two types.
The first type recommends or orders all vehicles within specific road sections to change their maneuvers~\citep{Hegyi2008,Carlson2010a,VandeWeg2014}.
The second type manipulates only a single or a certain percentage of vehicles~\citep{Nishi2013,Jerath2015,Han2017,Stern2018}. The maneuvers of the other vehicles are indirectly controlled by the manipulated vehicles.
Because moving the positions of infrastructures or adding them for improving the first type is generally expensive, the second type is expected to realize more flexible execution than the first type.
Therefore, this paper focuses on the second type.
In particular, seeking the manipulation of the maneuvers of as few vehicles as possible is challenging and will enhance the robustness of the second type of manipulation. Accordingly, this paper focuses on the manipulation of a single vehicle's maneuvers~\citep{Nishi2013,Stern2018} among the latter form of manipulation.

As a manipulation of a single vehicle's maneuvers, this paper addresses jam-absorption driving (JAD)~\citep{Nishi2013}, which removes a wide moving jam (a traffic jam whose downstream head and upstream tail move in the upstream direction~\citep{Kerner2004Physics}). JAD is composed of two consecutive actions: \textit{slow-in} and \textit{fast-out}. In the slow-in phase, a single vehicle (hereafter called the \textit{absorbing vehicle}) decelerates and maintains a low velocity to avoid being captured by a wide moving jam, as shown in Fig.~\ref{fig:schematic_view_JAD}(a). Because the supply of vehicles to the jam is cut off, the jam shrinks and finally disappears. After the jam has dissipated, the fast-out phase begins, and the absorbing vehicle promptly returns to following the vehicle just ahead of it.

One of the critical problems of JAD is the occurrence of secondary jams. The absorbing vehicle causes perturbations that propagate back upstream. These perturbations may grow into secondary jams because of the instability of traffic flows~\citep{Helbing2001,Treiber2013}, as shown in Fig.~\ref{fig:schematic_view_JAD}(b).
One possible way to restrict secondary jams is improving the instability of the traffic flow upstream of the absorbing vehicle by deploying connected and/or automated vehicles in this upstream flow. For instance, only a single CAV can stabilize a platoon of vehicles based on stability margin~\citep{Wang2018b}.
However, we do not set connected or automated vehicles upstream of the absorbing vehicle in this paper. As a baseline case, we assume that all vehicles upstream of the absorbing vehicle are human-driven vehicles (HDVs)~\citep{Taniguchi2015, He2017}.
The occurrence of secondary jams may be affected by the types of systems. Theoretical studies showed that a single connected and/or automated vehicle can stabilize entire traffic flows in ring roads~\citep{Wu2018,Zheng2018}. Hence, secondary jams are expected to be restricted by a single vehicle in ring roads. However, these theoretical results are limited to ring roads. We treat non-periodic roads in this paper.
Accordingly, we focus on the systems in which all vehicles upstream of the absorbing vehicle are HDVs and roads are not periodic. Although numerical simulations with microscopic car-following models investigated the occurrence of secondary jams in these systems~\citep{Taniguchi2015, He2017}, there is no theoretical condition to restrict secondary jams. Thus, we require some theoretical support to suppress these secondary jams.

We now specify the type of traffic jam targeted in this paper. JAD and other strategies have targeted various traffic jams, such as single wide moving jams on a road without bottlenecks~\citep{Popov2008, Hegyi2008, Nishi2013, Taniguchi2015, Wang2016JITS, Wang2016IEEE, He2017}, traffic jams whose downstream heads are fixed at a single bottleneck~\citep{Chen2014, Han2017, Ghiasi2017}, multiple wide moving jams occurring from a bottleneck~\citep{He2017}, and traffic jams on highway networks with many on- and off-ramps~\citep{Carlson2010b}.
Easing traffic jams on a large scale~\citep{Carlson2010b} drastically improves the flow of highway traffic. Additionally, because most traffic jams on highways are caused by bottlenecks, easing traffic jams fixed at bottlenecks would provide more benefits than easing wide moving jams.
Nevertheless, wide moving jams also deteriorate flow rates. The flow rate out of a wide moving jam was reported to be two-thirds of the free flow capacity~\citep{Kerner2004Physics}. Therefore, removing wide moving jams would also contribute to the realization of more efficient highway traffic. For instance, removing wide moving jams was shown to improve the total travel time by $35$ vehicle hours (veh-h) in field tests on real highways~\citep{Hegyi2010} and by $87.8\,$veh-h and more in numerical simulations~\citep{Wang2016JITS}.
Experiments to dissolve a wide moving jam using a real autonomous vehicle on a ring road also improved the flow rate by $14.1\,\%$ and fuel consumption by $39.8\,\%$~\citep{Stern2018}.
Accordingly, this paper mainly focuses on removing a single wide moving jam. We cause this traffic jam by inserting an initial perturbation into the traffic flow~\citep{Taniguchi2015, Wang2016JITS, Wang2016IEEE, He2017}.
In the later parts of this paper, we also investigate JAD scenarios to remove a wide moving jam on a single-lane road with inflows from other lanes, and a traffic jam whose upstream head is fixed at a bottleneck.

Our aim is to construct a theoretical condition for restricting secondary jams when JAD removes a wide moving jam, under the systems in which only HDVs are placed upstream of the absorbing vehicle on non-periodic roads.
As a traffic flow, we set a platoon of vehicles obeying a microscopic car-following model containing the instability. The car-following model we use is the intelligent driver model (IDM), which is widely used for highway traffic~\citep{Treiber2000, Treiber2013}.
To construct this condition, we apply the linear string stability to a macroscopic spatiotemporal structure of JAD. The linear string stability is a criterion that determines the decay or growth of infinitely small perturbations propagating through a platoon of vehicles~\citep{Wilson2008b, Treiber2013}.
Although the real highway traffic does not contain an infinite number of vehicles, treating an infinite number of vehicles makes the theoretical analysis more tractable. Accordingly, we set the platoon to be composed of a leading vehicle and an infinite number of following vehicles.
In addition, real highways frequently contain multiple-lane roads and bottlenecks (such as sags, tunnels, on-ramps, merging sections, and weaving sections). Nevertheless, the condition developed for a simple road will provide the foundation for conditions in more complex scenarios. Therefore, as a type of road segment, we consider a single-lane road~\citep{Nishi2013,Taniguchi2015,Han2017,He2017}, and do not insert loops or bottlenecks into it.
After constructing this condition, we categorize the behavior of the semi-infinite platoon subjected to an initial perturbation into three cases based on this condition. In the first case, the perturbation decays and JAD is not necessary. In the second and third cases, the perturbation grows into a wide moving jam that is removed by JAD. Although JAD does not cause secondary jams in the second case, JAD is responsible for the occurrence of secondary jams in the third case.
We investigate the sensitivity of the behavior with respect to the parameters of the IDM and the initial velocity of the platoon.
In addition, we numerically check the validity of a finite system version of this condition to suppress secondary jams in finite systems.
After investigating the behavior of this single-lane system, we construct conditions for restricting secondary jams in other semi-infinite systems with inflows from other lanes or a bottleneck. We utilize the condition for the single-lane system for constructing these conditions. We also investigate the behaviors in these more complex systems.
By theoretically guaranteeing the restriction of secondary jams in JAD scenarios, this paper elucidates the influence of a single vehicle's maneuvers on the improvement of highway traffic flow.

The remainder of this paper is organized as follows.
In Sec.~\ref{sec:related_methods}, we review the studies on JAD and related topics.
In Sec.~\ref{sec:model}, we define the system, initial conditions, initiation of a wide moving jam, and JAD maneuvers.
In Sec.~\ref{sec:numerical_simulations}, we consider finite systems in preparation for handling a semi-infinite system. We present the results of numerical simulations of JAD on finite systems and confirm the occurrence of secondary jams.
Section~\ref{sec:theoretical_analysis} reviews the linear string stability for microscopic car-following models, describes a macroscopic view of JAD in the semi-infinite system, and presents the condition for suppressing secondary jams. We numerically check a finite version of this condition and investigate the influence of the parameters on the behavior of the semi-infinite system by using this condition.
We also construct conditions for suppressing secondary jams in other semi-infinite systems with inflows from other lanes or a bottleneck and investigate the behaviors of these systems.
Section~\ref{sec:discussion} presents the conclusions to this study and a discussion of the results.
%%%%%%%%%%%%%%%%%%%%%%%%%%%%%%%%%%%%%%%%%%%%%%%%%%%%%%%%%%%%%%%%%%%%%%%%%%%%%%%%%%%%%
\section{\label{sec:related_methods}Related work}
%%%%%%%%%%%%%%%%%%%%%%%%%%%%%%%%%%%%%%%%%%%%%%%%%%%%%%%%%%%%%%%%%%%%%%%%%%%%%%%%%%%%%
We now review the studies on JAD. The concept of JAD dates back to the idea of removing a stop-and-go wave or a traffic jam using a single vehicle: to enlarge a single vehicle's front inter-vehicular distance in advance, and preventing the vehicle from being captured by the stop-and-go wave or the traffic jam~\citep{Beaty1998,Washino2003}.
This idea was formulated through the concept of a single pace car removing a traffic jam fixed at a bottleneck~\citep{Behl2010}. However, the formulation was limited to the timing of the pace car and did not consider the propagation of perturbations caused by it.
Later, JAD was defined as a slow-in and fast-out driving strategy for removing a wide moving jam performed by a single vehicle, and a theoretical framework of JAD was constructed with a microscopic traffic model~\citep{Nishi2013}.
Ref.~\citep{Nishi2013} also analyzed the propagation of perturbations caused by JAD, but as this model did not incorporate any instabilities, they were not able to analyze secondary jams induced by the instabilities.
The occurrence of secondary jams was investigated through numerical simulations with car-following models containing instabilities~\citep{Taniguchi2015}.
Ref.~\citep{He2017} also numerically investigated the occurrence of secondary jams using a non-parametric car-following model. Moreover, Ref.~\citep{He2017} proposed a two-step procedure of JAD and a method of selecting the absorbing vehicle, and numerically investigated the mitigation of multiple wide moving jams arising from a single bottleneck.
Besides, a wavelet transform was introduced into JAD to predict accurately the targeted traffic jam~\citep{Zheng2018a}.
A JAD experiment was conducted using five real vehicles in a closed environment, albeit without investigating secondary jams~\citep{Taniguchi2015TGF}.

There are many topics related to JAD, such as improving the traffic flow by changing the spatiotemporal maneuvers of vehicles rather than enhancing microscopic car-following performances.
First, we review the studies that utilize the maneuvers of a single vehicle or a certain percentage of vehicles. Researchers have studied on removing a wide moving jam or stabilizing the whole system by putting a single autonomous vehicle into ring roads.
An experiment using more than 20 real vehicles on a ring road demonstrated that a single autonomous vehicle's maneuvers can remove a wide moving jam, and improve total fuel consumption and flow rate~\citep{Stern2018} as well as emissions~\citep{Stern2019}.
A wide moving jam was numerically removed on a ring road in a similar way in numerical simulations with reinforcement learning~\citep{Wu2017}.
Ref.~\citep{Zheng2018} proved that a ring system is stabilizable by controlling only a single vehicle.
Researchers have also studied on removing or mitigating traffic jams on non-periodic roads.
A single vehicle's maneuvers removed a traffic jam fixed at a bottleneck in numerical simulations~\citep{Ghiasi2017}.
Mitigation of traffic jams fixed at a bottleneck through the maneuvers of one or more vehicles was analyzed based on the shock wave theory~\citep{Han2017}. Ref.~\citep{Han2017} introduced the probabilistic capacity drop at the bottleneck and analyzed expected delay savings provided by the maneuvers.
A special decelerating-accelerating-decelerating-accelerating maneuvers of a small number of vehicles reduced the travel time for passing through a sag using numerical simulations~\citep{Goni-Ros2016a}.
A moving bottleneck (a low-speed and high-density region) produced by a single vehicle upstream of a bottleneck numerically reduced fuel consumption~\citep{Ramadan2017} and travel time~\citep{Piacentini2018a}.
A concept of the influential subspace was proposed, which is the spatial region in which CVs are able to improve the traffic flow through their maneuvers~\citep{Jerath2015}.
Traffic jam at a sag section was numerically mitigated through a dynamic VSL combined with a vehicle-to-vehicle (V2V) communication system~\citep{Morino2016,Watanabe2018a}.
Note that tuning a single vehicle's microscopic car-following parameters also improves the stability of entire ring systems~\citep{Wu2018} as well as vehicle platoons~\citep{Wang2018b}.

Second, we review the dynamical manipulations of the maneuvers of all vehicles within specific road sections.
A dynamic VSL algorithm for removing a wide moving jam was proposed based on the shock wave theory~\citep{Lighthill1955, Richards1956} and named the speed controlling algorithm using shockwave theory (SPECIALIST)~\citep{Hegyi2008}. SPECIALIST lowers the speed limit of vehicles upstream of the jam. This low speed limit makes the propagation of the upstream tail of the jam slower than that of the downstream head of the jam. Therefore, the jam shrinks and finally disappears.
SPECIALIST and JAD are common strategies for shrinking and removing a wide moving jam dynamically. There are two main differences between them.
First, SPECIALIST instructs all the vehicles within some road sections to decelerate, whereas JAD only seeks to control a single vehicle.
The second difference concerns the density of vehicles just upstream of the jam. The density is assumed to be maintained in SPECIALIST, whereas the absorbing vehicle in JAD produces a large vacant space, causing the density to become zero.
SPECIALIST has been tested on a real highway~\citep{Hegyi2009, Hegyi2010}, equipped with predictions of jam propagations~\citep{Hegyi2013}, and combined with a cooperative car-following control~\citep{Wang2016JITS}.
In addition to SPECIALIST, the mainstream traffic flow control (MTFC) blocks the capacity drop at a bottleneck by restricting the flow entering the bottleneck, and MTFC is mainly controlled by VSL and/or RM~\citep{Carlson2010a, Carlson2010b}.
A distributed control~\citep{Popov2008} and a model predictive control~\citep{Hegyi2005a, Han2017a} of VSL mitigated a wide moving jam using numerical simulations.
Mitigation of traffic jam fixed at a bottleneck by a dynamic VSL was analyzed based on the shock wave theory~\citep{Chen2014}.
VSL, RM, and CVs with penetration ratio of $100\,\%$ were combined for removing a wide moving jam~\citep{VandeWeg2014}.
Time-space trajectories of CAVs were optimized for removing a traffic jam in the research field of the trajectory planning~\citep{Li2019}. Ref.~\citep{Li2019} imposed a microscopic string stability conditions on these vehicles.

Finally, we review eco-driving as a related research field of JAD. Eco-driving is cost-effective on highways~\citep{Schwarzkopf1977} and arterials with a single~\citep{Trayford1984, Sanchez2006, Rakha2011} or multiple~\citep{Mandava2009, Asadi2011, Barth2011, Nunzio2016} traffic signals (interested readers are referred to a recent review~\citep{Vahidi2018}).
Some eco-driving algorithms prevent vehicles from entering queues caused by red traffic signals by taking into account of spatiotemporal propagations of the queues~\citep{Chen2014TRR, He2015TRC, Yang2017b}.
These eco-driving algorithms and JAD are common strategies for avoiding being captured by queues or traffic jams. Nevertheless, eco-driving aims to minimize certain costs (such as fuel consumptions or emissions) and does not always aim to remove queues. In contrast, JAD is always employed to remove traffic jams.

Among the aforementioned studies, secondary jams were considered by SPECIALIST~\citep{Hegyi2010} and JAD \citep{Taniguchi2015,He2017}.
However, none of the aforementioned studies proposed theoretical conditions for restricting secondary jams in non-periodic systems where all the vehicles upstream of the absorbing vehicle are HDVs and obey a car-following model containing instabilities. Therefore, we pursue our aim of determining such a condition in the remainder of this paper.
%%%%%%%%%%%%%%%%%%%%%%%%%%%%%%%%%%%%%%%%%%%%%%%%%%%%%%%%%%%%%%%%%%%%%
\section{\label{sec:model}Model}
%%%%%%%%%%%%%%%%%%%%%%%%%%%%%%%%%%%%%%%%%%%%%%%%%%%%%%%%%%%%%%%%%%%%%
\subsection{\label{subsec:system}System}
\begin{figure}[t]
\centering
\includegraphics[width=\hsize]{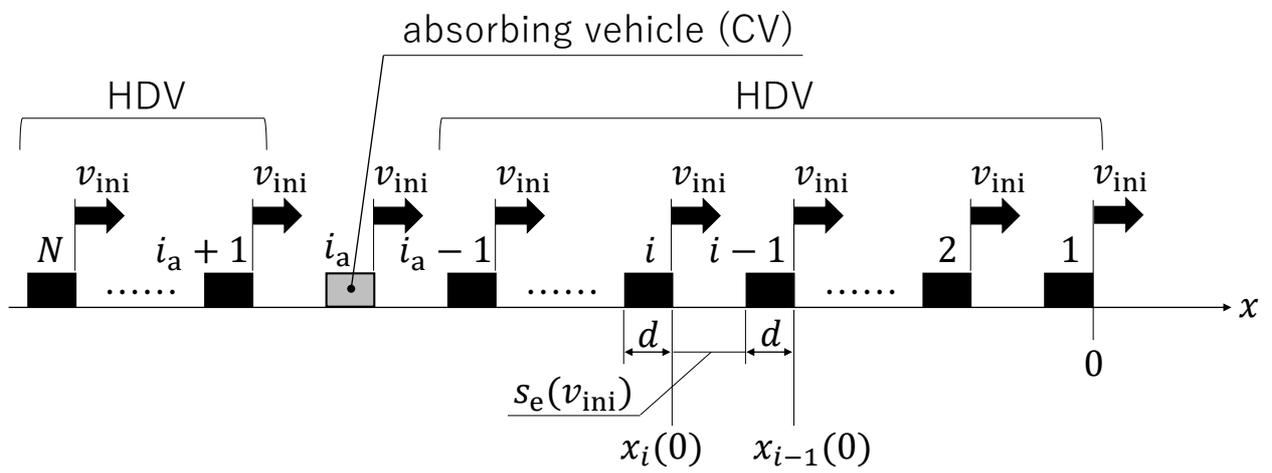}
\caption{Initial conditions of the system. All vehicles (vehicles $1, 2, \ldots, N$) run at velocity $v_{\rm ini}$ on a single-lane road without any loops or bottlenecks. All vehicles except for vehicle 1 have a front inter-vehicular distance of $s_{\rm e}(v_{\rm ini})$.
All vehicles except for the absorbing vehicle (vehicle $i_{\rm a}$) are HDVs. The absorbing vehicle is a CV.
If $N=\infty$, the system is semi-infinite. Otherwise, the system is finite.
}
\label{fig:ini_conds}
\end{figure}

%%%%%%%%%%%%%%%%%%%%%%%%%%%%%%%%%%%%%%%%%%%%%%%%%%%%%%%%%%%%%%%%%%%%%
We consider a platoon of vehicles on a single-lane road of infinite length without loops or bottlenecks, as shown in Fig.~\ref{fig:ini_conds}. Vehicle 1 is the leading vehicle of the platoon. Vehicle $i$ ($i = 1, \ldots, N$) is placed just behind vehicle $i-1$, where $N$ is the number of vehicles. Hence, vehicle $N$ is the last vehicle of the platoon.
We mainly use a semi-infinite platoon ($N=\infty$) in this paper, although finite platoons ($N$ is finite) are also considered.
$x_i(t)$ and $v_i(t)$ denote the position and velocity of vehicle $i$ at time $t$, respectively. $x_i(t)$ increases as vehicle $i$ moves from the upstream to a downstream position. All the vehicles move in the same direction (from upstream to downstream) and are banned from overtaking the vehicles in front of them.
We designate vehicle $i_{\rm a}$ $(2 \le i_{\rm a} \le N)$ as the absorbing vehicle.
We assume that all vehicles except for the absorbing vehicle are HDVs.
We assume that the absorbing vehicle is a CV and receives spatiotemporal information of the jam from infrastructures.

We define the car-following behavior of all vehicles except for two special vehicles: vehicle 1 and the absorbing vehicle. The movements of vehicle 1 and the absorbing vehicle are defined in Sec.~\ref{subsec:perturbation} and Sec.~\ref{subsec:JAD}, respectively.
Vehicle $i$ ($i \neq 1, i_{\rm a}$) follows vehicle $i-1$ and its acceleration is given by the IDM~\citep{Treiber2000, Treiber2013}:
\begin{align}
\dfrac{{\rm d} v_i(t)}{{\rm d} t} = a \left[ 1 - \left\{ \dfrac{v_i(t)}{v_0} \right\}^{\delta} -  \left\{ \dfrac{s^{\ast} \left( v_i(t), \Delta v_i(t) \right) }{s_i(t)} \right\}^2 \right],
\label{eq:IDM}
\end{align}
where $s^{\ast}(v_i(t), \Delta v_i(t))$ is defined as
\begin{align}
s^{\ast} \left( v_i(t), \Delta v_i(t) \right) = s_0 + \max \left\{ 0, T v_i(t) + \dfrac{v_i(t) \Delta v_i(t)}{2 \sqrt{ab}} \right\}.
\label{eq:s_asterisk}
\end{align}
$\Delta v_i(t)$ is the relative velocity between vehicles $i$ and $i-1$, and is given by
\begin{align}
\Delta v_i(t) = v_i(t) - v_{i-1}(t).
\label{eq:Delta_v_i_t}
\end{align}
$s_{i}(t)$ is the gap between the rear end of vehicle $i-1$ and the front end of vehicle $i$, which is written as
\begin{align}
s_{i}(t) = x_{i-1}(t) - x_i(t) - d.
\label{eq:s_i_t}
\end{align}
In the IDM, $a$ is the maximal acceleration, $b$ is a comfortable deceleration, $d$ is the length of each vehicle, $s_0$ is the gap in the halting state, $v_0$ is the desired velocity, which is not realized as long as $s_i(t)$ is finite, $T$ is the safe time gap, and $\delta$ is the exponent.
For simplicity, we assume that all vehicles have the same parameter values. We set these parameters to have typical values for highway traffic~\citep{Treiber2013}, as listed in Table~\ref{tab:param}.

In the IDM, the equilibrium gap $s_{\rm e}(v)$ is a function of velocity $v$. Because the equilibrium state is realized when $v_i(t) = v_{i-1}(t) = v$ and ${\rm d}v_i(t)/{\rm d}t = 0$ in Eq.~(\ref{eq:IDM}), $s_{\rm e}(v)$ is given by~\citep{Treiber2000,Treiber2013}:
\begin{align}
s_{\rm e}(v) = \dfrac{s_0 + v T}{\sqrt{1 - \left( \dfrac{v}{v_0} \right)^\delta}}.
\label{eq:s_e}
\end{align}
The IDM also has the equilibrium velocity $v_{\rm e}(s)$ that is the inverse function of $s_{\rm e}(v)$~\citep{Treiber2013}.

The IDM is a microscopic car-following model in which the acceleration of vehicle $i$ is given by a function $\tilde{a}_{\rm mic}$ of its velocity $v_i(t)$, gap $s_{i}(t)$, and relative velocity $\Delta v_i(t)$~\citep{Wilson2008b, Treiber2013}:
\begin{align}
\dfrac{{\rm d} v_i(t)}{{\rm d} t} = \tilde{a}_{\rm mic}(s_i(t), v_i(t), \Delta v_i(t)).
\label{eq:a_mic}
\end{align}

We set the initial conditions as shown in Fig.~\ref{fig:ini_conds}. At the initial time $t=0\,{\rm s}$, all the vehicles have initial velocity $v_{\rm ini}$, that is, $v_i(0)=v_{\rm ini}$ ($i = 1, \ldots, N$), and the initial position of vehicle 1 is zero, that is, $x_1(0)=0\,{\rm m}$. We place the following vehicles $i$ ($i = 2, \ldots, N$) at regular intervals:
\begin{align}
x_i(0) = x_{i-1}(0) - d - s_{\rm e}(v_{\rm ini}).
\label{eq:x_i_0}
\end{align}
Accordingly, the acceleration of vehicle $i$ ($i = 2, \ldots, N$) is initially zero. Note that $\lim_{v \to v_0 - 0} s_{\rm e}(v)=\infty$ and it is not possible to set a homogeneous platoon of vehicles in equilibrium with a velocity of $v_0$. Therefore, we set $v_{\rm ini} < v_0$.
%%%%%%%%%%%%%%%%%%%%%%%%%%%%%%%%%%%%%%%%%%%%%%%%%%%%%%%%%%%%%%%%%%%%%%%%%%%%%
\subsection{\label{subsec:perturbation}Producing a wide moving jam}
%%%%%%%%%%%%%%%%%%%%%%%%%%%%%%%%%%%%%%%%%%%%%%%%%%%%%%%%%%%%%%%%%%%%%%%%%%%%%
As a target jam to be removed, we produce a wide moving jam by imposing a perturbation on the traffic flow~\citep{Taniguchi2015, Wang2016JITS, Wang2016IEEE, He2017}. We cause this perturbation through four consecutive actions of vehicle 1~\citep{Taniguchi2015}.
\begin{description}
\item[(i)] At the initial time $t=0\,{\rm s}$, vehicle 1 starts to decelerate with acceleration $-\alpha_{\rm p}$ from its initial velocity $v_{\rm ini}$. Its velocity eventually becomes zero.
\item[(ii)] Vehicle 1 stops for a period of $T_{\rm p}$.
\item[(iii)] Vehicle 1 then starts to accelerate with acceleration $\alpha_{\rm p}$. Its velocity eventually returns to $v_{\rm ini}$.
\item[(iv)] Finally, vehicle 1 maintains its velocity at $v_{\rm ini}$.
\end{description}
We set $\alpha_{\rm p}=1\,{\rm m}/{\rm s}^2$ and $T_{\rm p}=1\,{\rm s}$.

The perturbation grows into a single wide moving jam if its amplitude is sufficiently large~\citep{Kerner2004Physics, Helbing2001}. As long as we do not perform JAD, the jam propagates to the last vehicle. We define $v_{\rm R}$ as the velocity of the downstream head of the jam (the velocity of the rarefaction wave) and $v_{\rm S}$ as the velocity of the upstream tail of the jam (the velocity of the shock wave). After a sufficiently long time from the birth of the jam, the jam captures vehicles at a constant time interval and discharges them at another constant time interval. Hence, $v_{\rm R}$ and $v_{\rm S}$ eventually become constant values.
The growth or decay of the jam is characterized by the relationship between $v_{\rm R}$ and $v_{\rm S}$. If $v_{\rm S} < v_{\rm R} < 0$, the length of the jam will increase~\citep{Nishi2013, Taniguchi2015, Wang2016JITS, Wang2016IEEE, Han2017a}. If $v_{\rm S} = v_{\rm R} < 0$, its length will remain constant~\citep{Popov2008, He2017}. If $v_{\rm R} < v_{\rm S} < 0$, its length will decrease~\citep{Jerath2015}.
%%%%%%%%%%%%%%%%%%%%%%%%%%%%%%%%%%%%%%%%%%%%%%%%%%%%%%%%%%%%%%%%%%%%%
\subsection{\label{subsec:JAD}Jam-absorption driving}
%%%%%%%%%%%%%%%%%%%%%%%%%%%%%%%%%%%%%%%%%%%%%%%%%%%%%%%%%%%%%%%%%%%%%
We assume that a wide moving jam occurs following vehicle 1's perturbation and that its length is constant or increasing ($v_{\rm S} \le v_{\rm R} < 0$). We remove the jam using JAD. The JAD process employs the following three steps similar to the process used in Ref.~\citep{Taniguchi2015}.
\begin{description}
\item[(i)] In the slow-in phase, the absorbing vehicle (vehicle $i_{\rm a}$) starts decelerating with a constant acceleration $-\alpha_{\rm a}$ upon the occurrence of the perturbation, that is, at $t=0\,{\rm s}$.
\item[(ii)] After its velocity becomes $v_{\rm a} (< v_{\rm ini})$, vehicle $i_{\rm a}$ stops decelerating and maintains its velocity at $v_{\rm a}$ for a period $T_{\rm a}$. Its motion produces a vacant space upstream of the jam and stops the supply of vehicles to the jam. Therefore, the jam shrinks and finally disappears.
\item[(iii)] After running at $v_{\rm a}$ for $T_{\rm a}$, vehicle $i_{\rm a}$ starts following the vehicle just ahead of it according to the IDM as the fast-out phase.
\end{description}
Hereafter, we name $v_{\rm a}$ the \textit{absorbing velocity}. We set $\alpha_{\rm a} = 1\,{\rm m}/{\rm s}^2$.

Note that detecting traffic jams and estimating their propagations in real time~\citep{Hegyi2013} are beyond the scope of this paper. Instead, we assume that the absorbing vehicle knows the spatiotemporal information of the jam (such as the position of the downstream head of the jam as a function of time) at the initial time.
Selecting the appropriate absorbing vehicle~\citep{He2017} is also beyond the scope of this paper. Instead, we choose the absorbing vehicle from the beginning of a run.
These simplifications do not detract from the essence of the problem of secondary jams.

To set $v_{\rm a}$ and $T_{\rm a}$ for a finite $N$, we perform a run in which JAD is not activated in advance. Once a wide moving jam has been produced, it propagates upstream. After this run is complete, we obtain the time and position of the vehicle just ahead of the absorbing vehicle (that is, vehicle $i_{\rm a} - 1$) when it escapes from the jam. We name these values $t_{i_{\rm a} - 1}^{\rm R}$ and $x_{i_{\rm a} - 1}^{\rm R}$, respectively. We judge that vehicles escape from the jam when their velocity becomes greater than $1\,{\rm m}/{\rm s}$. We use $t_{i_{\rm a} - 1}^{\rm R}$ and $x_{i_{\rm a} - 1}^{\rm R}$ to obtain $v_{\rm a}$ and $T_{\rm a}$.

Next, we perform a second run in which JAD is activated.
We set the absorbing vehicle to reach position $x_{i_{\rm a} - 1}^{\rm R} - X_{\rm buf}$ at time $t = t_{i_{\rm a} - 1}^{\rm R} + T_{\rm buf}$. $T_{\rm buf}$ and $X_{\rm buf}$ are temporal and spatial buffers, respectively, and we set $T_{\rm buf} = 10\,{\rm s}$ and $X_{\rm buf} = 100\,{\rm m}$.
In addition, we set $t = t_{i_{\rm a} - 1}^{\rm R} + T_{\rm buf}$ to be the time when a period $T_{\rm a}$ has passed since the absorbing vehicle's velocity becomes $v_{\rm a}$.
The value of $v_{\rm a}$ satisfying these settings is the solution of the quadratic equation:
\begin{align}
v_{\rm a}^2 + 2 c_1 v_{\rm a} - c_2 = 0,
\label{eq:v_a_quadratic_eq}
\end{align}
where
\begin{align}
c_1 = \alpha_{\rm a} (t_{i_{\rm a} - 1}^{\rm R} + T_{\rm buf}) - v_{\rm ini}, \ \ \ \
c_2 = 2 \alpha_{\rm a} \left\{x_{i_{\rm a} - 1}^{\rm R} - X_{\rm buf} - x_{i_{\rm a}}(0)\right\} - v_{\rm ini}^2.
\label{eq:c_1_c_2}
\end{align}
Because we set $\alpha_{\rm a} = 1\,{\rm m}/{\rm s}^2$, $c_1$ and $c_2$ are usually positive for large $i_{\rm a}$.
Therefore, we obtain $v_{\rm a}$ as follows:
\begin{align}
v_{\rm a} = \sqrt{c_1^2 + c_2} - c_1 = \dfrac{c_2}{c_1 + \sqrt{c_1^2 + c_2}}.
\label{eq:v_a_JAD_simulation}
\end{align}
We obtain $T_{\rm a}$ using $v_{\rm a}$ as a parameter:
\begin{align}
T_{\rm a} = t_{i_{\rm a} - 1}^{\rm R} + T_{\rm buf} - \dfrac{v_{\rm ini} - v_{\rm a}}{\alpha_{\rm a}}.
\label{eq:T_a_JAD_simulation}
\end{align}
%%%%%%%%%%%%%%%%%%%%%%%%%%%%%%%%%%%%%%%%%%%%%%%%%%%%%%%%%%%%%%%%%%%%%%%%%%%%%%%%%%%%%%%%
\section{\label{sec:numerical_simulations}Numerical simulations of secondary jams}
%%%%%%%%%%%%%%%%%%%%%%%%%%%%%%%%%%%%%%%%%%%%%%%%%%%%%%%%%%%%%%%%%%%%%%%%%%%%%%%%%%%%%%%%
As preparation for our theoretical treatment of secondary jams in the semi-infinite system ($N=\infty$), we consider secondary jams in finite systems using numerical simulations.
The occurrence of secondary jams in removing a single wide moving jam was investigated using $10^3$ vehicles~\citep{Taniguchi2015} or fewer than $10^2$ vehicles~\citep{He2017}.
Because large values of $N$ enable us to check the validity of the theoretical condition, we set the number of vehicles to $N \in \left\{ 10^3, 10^4, 10^5 \right\}$.

In our numerical simulations, we vary $v_{\rm ini}$ from $20.5$--$26.0\,{\rm m}/{\rm s}$ in increments of $0.5\,{\rm m}/{\rm s}$. The IDM parameters are listed in Table~\ref{tab:param}.
We set $i_{\rm a} = 2 N / 5 + 1$, so that $i_{\rm a}$ is approximately proportional to $N$.
We update $t$ from $0\,{\rm s}$ to $\,2 N\,{\rm s}$ at regular time intervals of $0.1\,{\rm s}$ in each run. This maximal time is sufficiently large to check the occurrence of secondary jams. We use the ballistic method~\citep{Treiber2015} as a numerical integration scheme. We calculate the exact position and velocity of vehicle 1 from $t = 0\,{\rm s}$ to $t = 2 N\,{\rm s}$, and the exact position and velocity of the absorbing vehicle from $t = 0\,{\rm s}$ to $t = t_{i_{\rm a} - 1}^{\rm R} + T_{\rm buf}$.
In each run, vehicle 1's perturbation grows into a wide moving jam and the absorbing vehicle removes it through JAD. After completing a run, we detect the occurrence of secondary jams as follows. If $v_{N}(t)$ falls below $1\,{\rm m}/{\rm s}$ at any point in the run, then at least one secondary jam has occurred. Otherwise, no secondary jams have occurred.

\begin{figure}[t]
\centering
\includegraphics[width=\hsize]{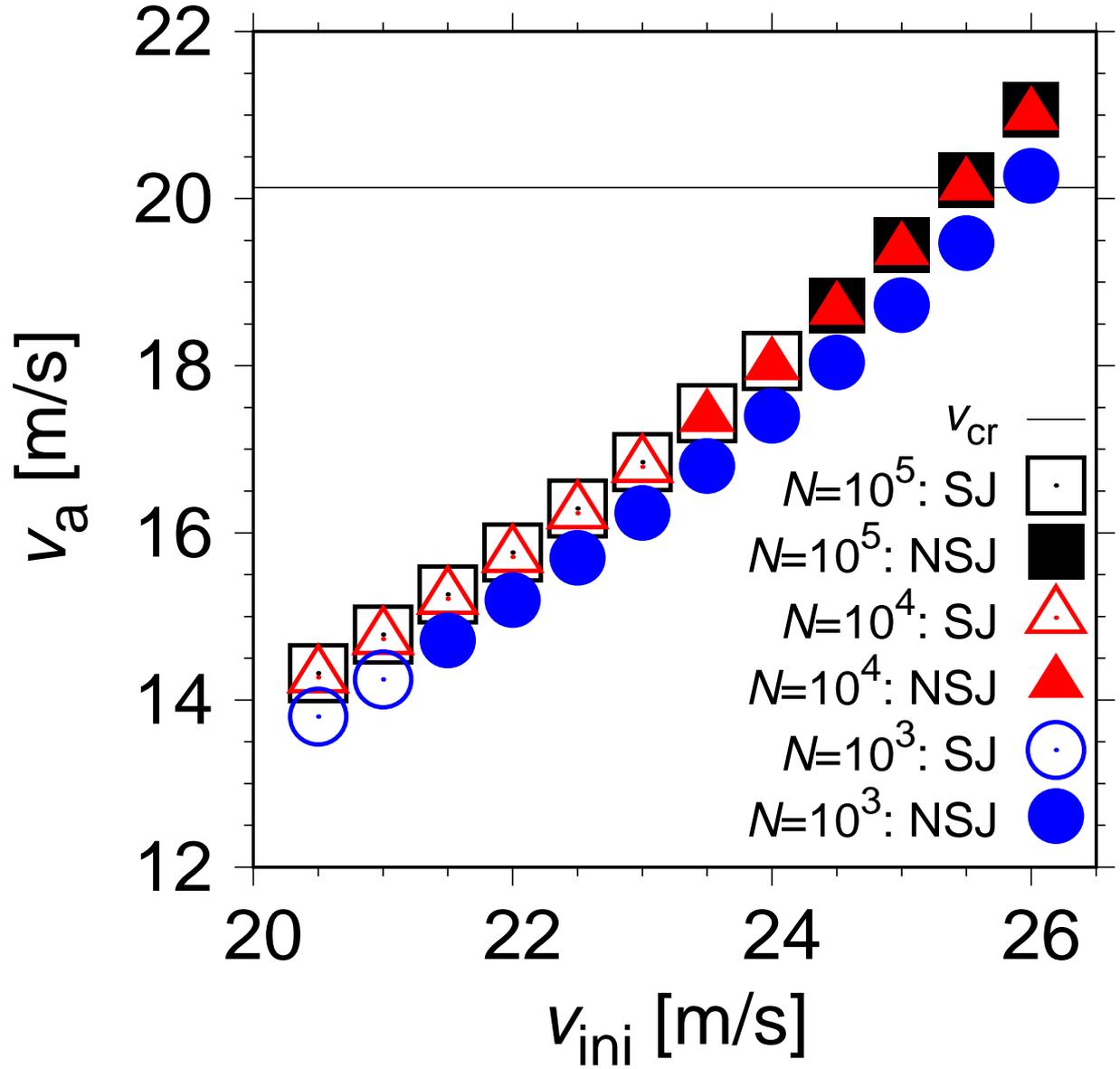}
\caption{
$v_{\rm a}$ as a function of $v_{\rm ini}$ obtained from numerical simulations.
In the numerical simulations, we set $i_{\rm a} = 2 N / 5 +1$ and $N\in\left\{10^3, 10^4, 10^5\right\}$.
The parameters of the IDM are given in Table~\ref{tab:param}.
The open and filled symbols represent the cases in which secondary jams (SJ) and no secondary jams (NSJ) occur throughout a run, respectively.
The thin horizontal line denotes $v_{\rm cr} = 20.13\,{\rm m}/{\rm s}$.
}
\label{fig:vini_va_secondary_jam}
\end{figure}

Figure~\ref{fig:vini_va_secondary_jam} shows $v_{\rm a}$ as a function of $v_{\rm ini}$ obtained from numerical simulations under $i_{\rm a} = 2 N / 5 + 1$ and $N \in \left\{ 10^3, 10^4, 10^5 \right\}$. The open/filled symbols denote whether secondary jams did/did not occur, respectively.
For each value of $i_{\rm a}$, $v_{\rm a}$ increases as $v_{\rm ini}$ increases. Secondary jams are less likely to occur as $v_{\rm a}$ becomes larger, as reported by numerical simulations~\citep{Taniguchi2015, He2017}.
The threshold value of $v_{\rm a}$ that determines whether secondary jams occur becomes larger as $i_{\rm a}$ increases.

\begin{figure}[t]
\centering
\includegraphics[width=\hsize]{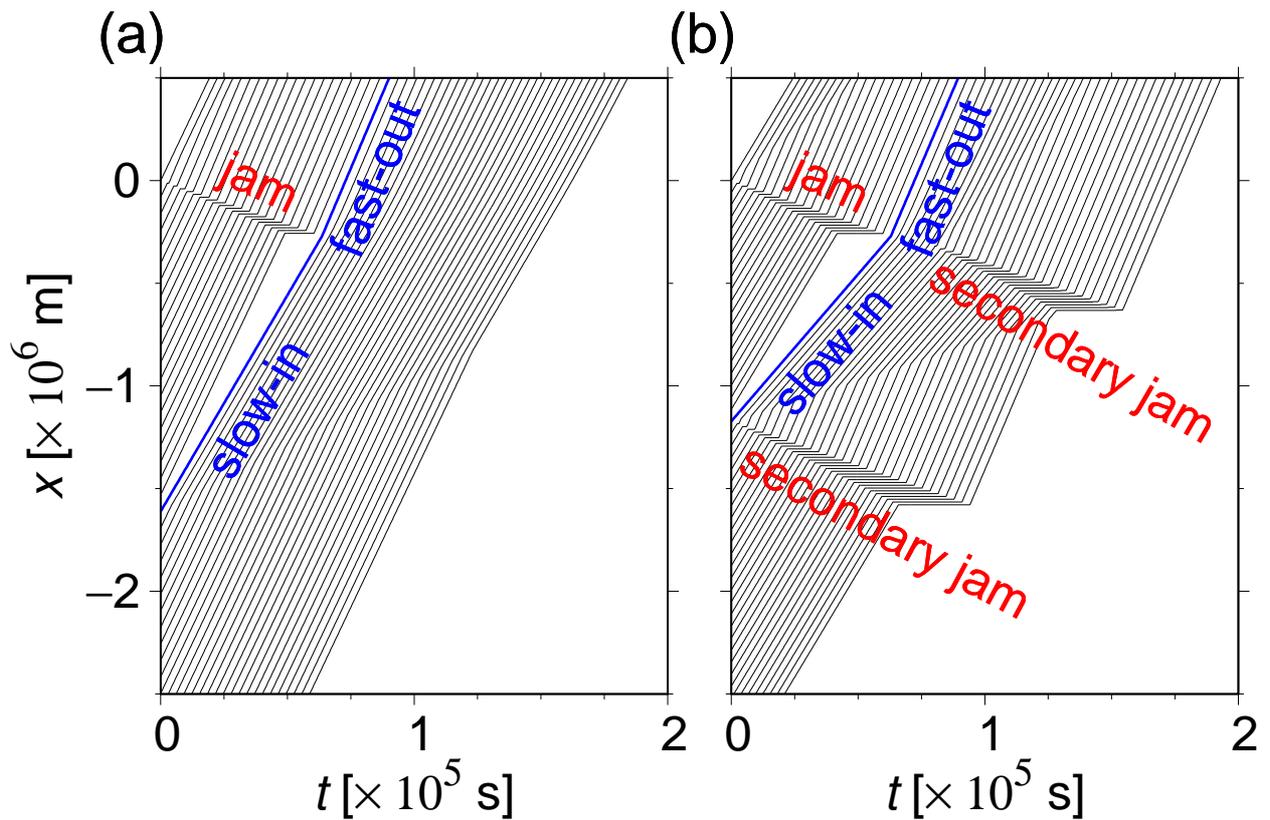}
\caption{Examples of the time-space diagrams of JAD.
A single vehicle (called the absorbing vehicle) removes a wide moving jam by two actions: slow-in and fast-out. In the slow-in phase, the absorbing vehicle decelerates from $v_{\rm ini}$ to the absorbing velocity $v_{\rm a}$ and maintains $v_{\rm a}$. In the fast-out phase, it returns to following the vehicle just ahead of it. We set $N=10^5$ and plot the tracks of vehicles $1, 2001, 4001, \ldots, 98001$ and $10^5$. The absorbing vehicle is vehicle $40001$ $(i_{\rm a} = 40001)$ corresponding to the thick blue lines.
(a) Initial velocity $v_{\rm ini}=26.0$ ${\rm m}/{\rm s}$. The absorbing vehicle does not cause secondary jams. (b) $v_{\rm ini}=20.5$ ${\rm m}/{\rm s}$. Two secondary jams occur.
}
\label{fig:schematic_view_JAD}
\end{figure}

Figure~\ref{fig:schematic_view_JAD} shows two time-space diagrams under $N=10^5$ and two different values of $v_{\rm ini}$. No secondary jam occurred under $v_{\rm ini} = 26.0\,{\rm m}/{\rm s}$, as shown in Fig.~\ref{fig:schematic_view_JAD}(a), whereas two secondary jams occurred under $v_{\rm ini} = 20.5\,{\rm m}/{\rm s}$ because of the slow-in and fast-out behavior, as shown in Fig.~\ref{fig:schematic_view_JAD}(b).
Note that after vehicles escape from the wide moving jam, they run with a velocity higher than $v_{\rm ini}$ for a certain period, and then decelerate toward vehicle 1's final velocity $v_{\rm ini}$. Therefore, $v_{\rm R}$ is slower than $v_{\rm S}$.
%%%%%%%%%%%%%%%%%%%%%%%%%%%%%%%%%%%%%%%%%%%%%%%%%%%%%%%%%%%%%%%%%%%%%%
\section{\label{sec:theoretical_analysis}Theoretical Analysis}
%%%%%%%%%%%%%%%%%%%%%%%%%%%%%%%%%%%%%%%%%%%%%%%%%%%%%%%%%%%%%%%%%%%%%%
Throughout of this section, we treat semi-infinite systems in which both $i_{\rm a}$ and $N - i_{\rm a}$ (the number of vehicles upstream of the absorbing vehicle) are infinite.
%%%%%%%%%%%%%%%%%%%%%%%%%%%%%%%%%%%%%%%%%%%%%%%%%%%%%%%%%%%%%%%%%%%%%%%%%%%%%%%%%%%%%%%%%%%%%%%%%%%%%%%%%%%%%%%%%%%%%%%
\subsection{\label{subsec:lss_necessity}Linear string stability for theoretically treating secondary jams}
%%%%%%%%%%%%%%%%%%%%%%%%%%%%%%%%%%%%%%%%%%%%%%%%%%%%%%%%%%%%%%%%%%%%%%%%%%%%%%%%%%%%%%%%%%%%%%%%%%%%%%%%%%%%%%%%%%%%%%%
To analyze the occurrence of secondary jams, we focus on the stability of a platoon composed of the absorbing vehicle (vehicle $i_{\rm a}$) and those following it (that is, vehicles $i_{\rm a}, i_{\rm a} + 1, i_{\rm a} + 2, \ldots$). If this platoon is unstable against the absorbing vehicle's perturbations (such as deceleration from velocity $v_{\rm ini}$ to $v_{\rm a}$ or acceleration from velocity $v_{\rm a}$ to a higher velocity), perturbations grow and finally become secondary jams. If this platoon is stable against perturbations, the perturbations are restricted from growing into secondary jams.

The stability of a platoon of vehicles against perturbations has been established as the string stability (interested readers are referred to recent reviews~\citep{Dey2016, Li2018} and a recent book~\citep{Treiber2013}).
The string stability is defined as follows~\citep{Treiber2013}: ``Traffic flow is string stable if local perturbations decay \textit{everywhere} even in \textit{arbitrarily long} vehicle platoons."
The stability is categorized into linear stability (stability against infinitesimal perturbations) and nonlinear stability (stability against finite perturbations)~\citep{Helbing2001,Treiber2013}.
In general, the analysis of linear stability (for instance, Ref.~\citep{Bando1995}) is more tractable than that of nonlinear stability~\citep{Helbing2009EPJB}. Therefore, we use linear string stability to determine the occurrence of secondary jams.

We briefly review the condition of linear string stability for the general microscopic car-following model given by Eq.~(\ref{eq:a_mic})~\citep{Wilson2008b, Treiber2013}. Let us consider a platoon in equilibrium composed of an infinite number of vehicles. Each vehicle obeys this model and has the equilibrium inter-vehicular distance $s_{\rm e}$, equilibrium velocity $v_{\rm e}(s_{\rm e})$ and relative velocity of zero. The condition for this platoon to have linear string stability is given by~\citep{Wilson2008b, Treiber2013}:
\begin{align}
\dfrac{{\rm d} v_{\rm e}(s_{\rm e})}{{\rm d} s} \le
- \dfrac{1}{2} \dfrac{\partial \tilde{a}_{\rm mic}(s_{\rm e}, v_{\rm e}, 0)}{\partial v_i}
- \dfrac{\partial \tilde{a}_{\rm mic}(s_{\rm e}, v_{\rm e}, 0)}{\partial \Delta v_i}.
\label{eq:string_stability_a_mic}
\end{align}
Condition~(\ref{eq:string_stability_a_mic}) is only applicable in the models satisfying
\begin{align}
\dfrac{\partial \tilde{a}_{\rm mic}(s_{\rm e}, v_{\rm e}, 0)}{\partial v_i} < 0
\label{eq:string_stability_a_mic_necessary_cond_1}
\end{align}
and
\begin{align}
\dfrac{{\rm d} v_{\rm e}(s_{\rm e})}{{\rm d} s} \ge 0.
\label{eq:string_stability_a_mic_necessary_cond_2}
\end{align}
Ordinarily, microscopic car-following models including the IDM satisfy these two conditions.

\begin{figure}[t]
\centering
\includegraphics[width=\hsize]{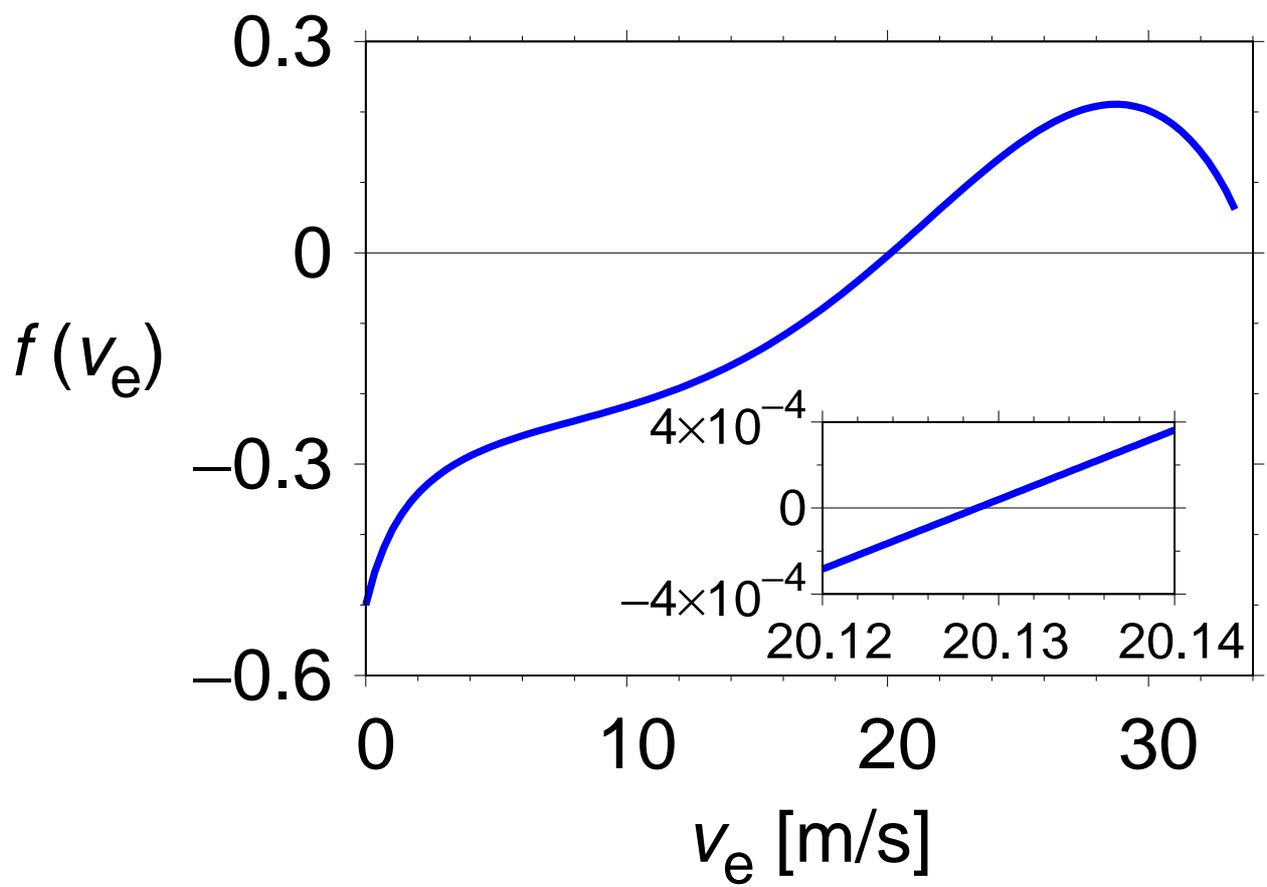}
\caption{$f(v_{\rm e})$ under the IDM parameter settings listed in Table~\ref{tab:param}. We plot the zero line as a visual guide. $v_{\rm cr}$ realizing $f(v_{\rm cr})=0$ is approximately $20.13\,{\rm m}/{\rm s}$.}
\label{fig:v_cr}
\end{figure}

We represent condition~(\ref{eq:string_stability_a_mic}) for the IDM as a condition of $v_{\rm e}$.
The three variables appearing in condition~(\ref{eq:string_stability_a_mic}) are given by
\begin{align}
\dfrac{{\rm d} v_{\rm e}(s_{\rm e})}{{\rm d} s} = \dfrac{\left\{ 1 - \left(\dfrac{v_{\rm e}}{v_0}\right)^{\delta}\right\}^{3/2}}{\dfrac{\delta s_0 v_{\rm e}^{\delta - 1}}{2 v_0^{\delta}} + T\left\{ 1 + \left(\dfrac{\delta}{2} - 1\right) \left(\dfrac{v_{\rm e}}{v_0}\right)^{\delta} \right\}},
\label{eq:v_e_s_e_prime}
\end{align}
\begin{align}
\dfrac{\partial \tilde{a}_{\rm mic}(s_{\rm e}, v_{\rm e}, 0)}{\partial v_i} = -a \left[ \dfrac{\delta v_{\rm e}^{\delta - 1}}{v_0^{\delta}} + \dfrac{2 T}{s_0 + v_{\rm e} T} \left\{ 1 - \left( \dfrac{v_{\rm e}}{v_0} \right)^{\delta} \right\} \right]
\label{eq:partial_a_mic_partial_v}
\end{align}
and
\begin{align}
\dfrac{\partial \tilde{a}_{\rm mic}(s_{\rm e}, v_{\rm e}, 0)}{\partial \Delta v_i} = - \dfrac{ v_{\rm e} }{ s_0 + v_{\rm e} T } \sqrt{ \dfrac{a}{b} } \left\{ 1 - \left( \dfrac{v_{\rm e}}{v_0} \right)^{\delta} \right\}.
\label{eq:partial_a_mic_partial_delta_v}
\end{align}
%%%%%%%%%%%%
Therefore, we rewrite condition~(\ref{eq:string_stability_a_mic}) as
\begin{align}
f(v_{\rm e}) \equiv a \left[ \dfrac{ \delta v_{\rm e}^{\delta - 1} }{ 2 v_0^{\delta} } + \dfrac{ 1 - \left( \dfrac{ v_{\rm e} }{ v_0 } \right)^{\delta} }{s_0 + v_{\rm e} T} \left( T + \dfrac{ v_{\rm e} }{ \sqrt{a b} } \right) \right] - \dfrac{ \left\{ 1 - \left( \dfrac{v_{\rm e}}{v_0} \right)^{\delta} \right\}^{3/2} }{ \dfrac{ \delta s_0 v_{\rm e}^{\delta - 1} }{ 2 v_0^{\delta} } + T \left\{ 1 + \left( \dfrac{\delta}{2} - 1 \right) \left( \dfrac{v_{\rm e}}{v_0} \right)^{\delta} \right\}} \ge 0,
\label{eq:string_stability_IDM}
\end{align}
where the function $f(v_{\rm e})$ is used for simplicity.
Figure~\ref{fig:v_cr} shows the shape of $f(v_{\rm e})$ under the parameter settings listed in Table~\ref{tab:param}. We define $v_{\rm cr}$ as the maximum value of $v_{\rm e}$ satisfying $f(v_{\rm cr})=0$ and $0 < v_{\rm cr} < v_0$. Figure~\ref{fig:v_cr} shows that $v_{\rm cr} = 20.13\,{\rm m}/{\rm s}$. Besides, $f(v_{\rm e})$ is negative for $0 \le v_{\rm e} < v_{\rm cr}$ and positive for $v_{\rm cr} < v_{\rm e} \le v_0$.
We rewrite condition~(\ref{eq:string_stability_IDM}) as
\begin{align}
v_{\rm cr} \le v_{\rm e} < v_0.
\label{eq:string_stability_IDM_v_cr}
\end{align}
Note that there may be more than two values of $v_{\rm e}$ for which $f(v_{\rm e})=0$ if the IDM parameters are set to other values. In this case, we focus on the uppermost region of $v_{\rm e}$ realizing $f(v_{\rm e}) \ge 0$ to determine $v_{\rm cr}$.

Note that the linear stability condition of the IDM is also given by~\citep{Treiber2013}:
\begin{align}
\left\{ \dfrac{{\rm d} v_{\rm e}(s_{\rm e})}{{\rm d} s} \right\}^2 \le
\dfrac{a(s_0 + v_{\rm e} T)}{s_{\rm e}^2} \left\{ \dfrac{s_0 + v_{\rm e} T}{s_{\rm e}} + \dfrac{v_{\rm e}}{\sqrt{ab}} \dfrac{{\rm d} v_{\rm e}(s_{\rm e})}{{\rm d} s} \right\}.
\label{eq:string_stability_IDM_Treiber_Kesting}
\end{align}
Because a function of $v_{\rm e}$ is easy to use for determining string stability in JAD, we use condition (\ref{eq:string_stability_IDM_v_cr}) instead of condition (\ref{eq:string_stability_IDM_Treiber_Kesting}) in this paper.
%%%%%%%%%%%%%%%%%%%%%%%%%%%%%%%%%%%%%%%%%%%%%%%%%%%%%%%%%%%%%%%%%%%%%%%%%%%%%%%%%%%%%
\subsection{\label{subsec:macro_view}Macroscopic time-space diagram of JAD}
\begin{figure}[t]
\centering
\includegraphics[width=\hsize]{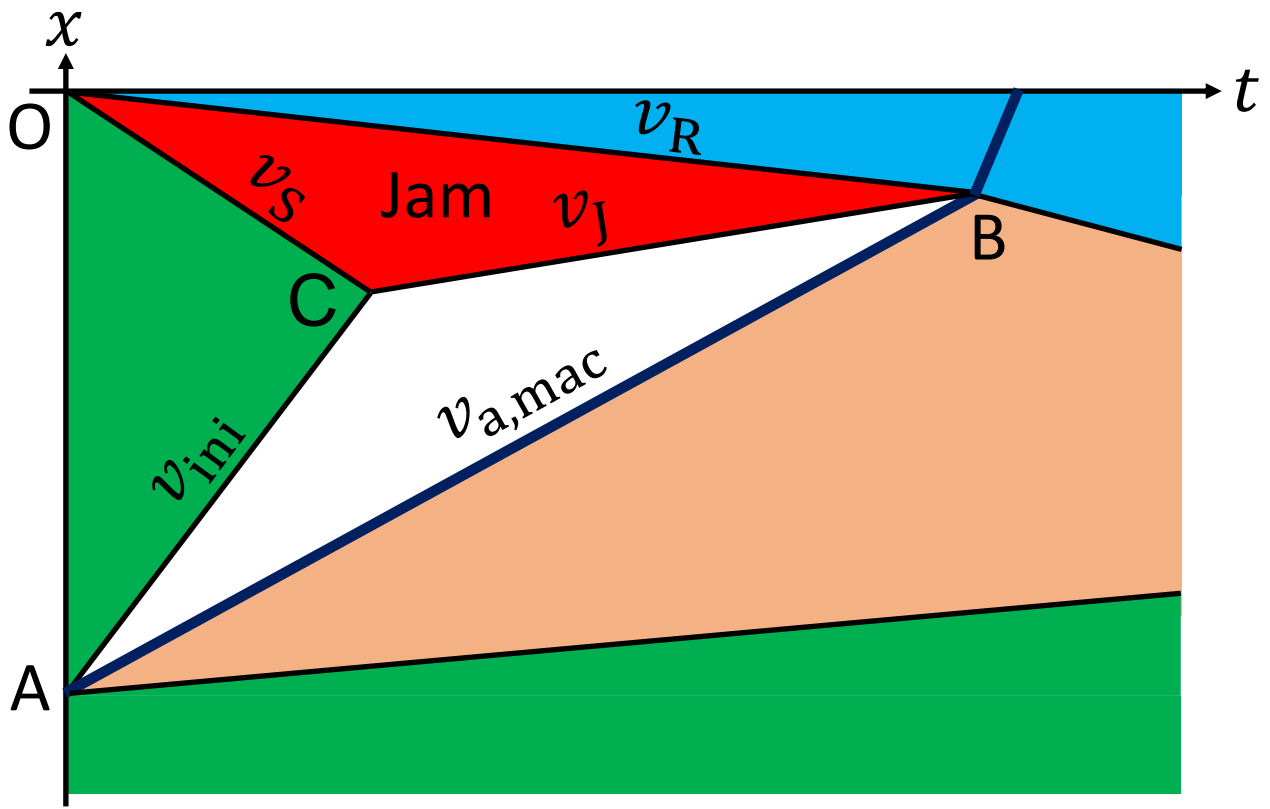}
\caption{Macroscopic view of JAD on a time-space diagram. We represent a wide moving jam by the triangle OBC. The absorbing vehicle does not pass through point C. It goes from point A to B at velocity $v_{\rm a, mac}$ during the slow-in phase. Its velocity increases at point B, signifying the fast-out phase.}
\label{fig:macro_view}
\end{figure}

%%%%%%%%%%%%%%%%%%%%%%%%%%%%%%%%%%%%%%%%%%%%%%%%%%%%%%%%%%%%%%%%%%%%%%%%%%%%%%%%%%%%%
To apply linear string stability to JAD, we simplify the time-space diagram of JAD to give the macroscopic time-space diagram shown in Fig.~\ref{fig:macro_view}.
In this macroscopic diagram, we ignore the time it takes for vehicles to decelerate or accelerate from one velocity to another. Therefore, we use only straight lines to depict the tracks of vehicles in the time-space diagram.
In addition, we ignore the period of the halting state $T_{\rm p}$, that is, vehicle 1 completes the chain of deceleration and acceleration instantly at the initial time $t=0$.
Moreover, we approximate a wide moving jam occurring at the origin O (0, 0) with initial length zero.
Furthermore, we approximate $v_{\rm R}$ and $v_{\rm S}$ as constants from the initial time.
These simplifications do not impair the essence of the time-space diagrams of JAD if $i_{\rm a}$ and $N-i_{\rm a}$ are sufficiently large.
Additionally, we assume that the jam propagates upstream and its length is constant or increasing (that is, $v_{\rm S} \le v_{\rm R} < 0$).

We define $v_{\rm a, mac}$ as the absorbing velocity in this macroscopic time-space diagram. The absorbing vehicle decelerates instantly from $v_{\rm ini}$ to $v_{\rm a, mac}$ at point A (0, $x_{\rm A}$), where $x_{\rm A} = x_{i_{\rm a}}(0) < 0$. It performs JAD and encounters the downstream head of the jam at point B ($t_{\rm B}$, $x_{\rm B}$), where its velocity changes from $v_{\rm a, mac}$ to a velocity higher than $v_{\rm ini}$. If it does not perform JAD, it will encounter the upstream tail of the jam at point C ($t_{\rm C}$, $x_{\rm C}$). $v_{\rm R}$ and $v_{\rm S}$ are given by the slope of the lines OB and OC, respectively. We define $v_{\rm J} (\ge 0)$ as the velocity of the vehicles inside the jam, which is given by the slope of the line BC.
The absorbing velocity $v_{\rm a, mac}$ is given by
\begin{align}
v_{\rm a, mac} = \dfrac{ (v_{\rm J} - v_{\rm R}) v_{\rm ini} + (v_{\rm R} - v_{\rm S}) v_{\rm J} }{ v_{\rm J} - v_{\rm S} }.
\label{eq:v_a_mac}
\end{align}
If $v_{\rm J}=0$, which is true in the case of the IDM, $v_{\rm a, mac}$ can be simplified to
\begin{align}
v_{\rm a, mac} = \dfrac{ v_{\rm R} }{ v_{\rm S} } v_{\rm ini}.
\label{eq:v_a_mac_simplified}
\end{align}
Finally, we note the relationship between $v_{\rm a, mac}$ and $v_{\rm a}$ given by Eq.~(\ref{eq:v_a_JAD_simulation}).
The time and position at which the wide moving jam arises are finite.
Those in which $v_{\rm S}$ and $v_{\rm R}$ become constant are also finite, as are those in which the absorbing vehicle's velocity becomes $v_{\rm a}$.
Therefore, as $i_{\rm a}$ goes to infinity, $v_{\rm a}$ converges to $v_{\rm a, mac}$.
%%%%%%%%%%%%%%%%%%%%%%%%%%%%%%%%%%%%%%%%%%%%%%%%%%%%%%%%%%%%%%%%%%%%%%%%%%%%%%%%%%%%%%%
\subsection{\label{subsec:lss_cond_JAD}Linear string stability condition of JAD}
%%%%%%%%%%%%%%%%%%%%%%%%%%%%%%%%%%%%%%%%%%%%%%%%%%%%%%%%%%%%%%%%%%%%%%%%%%%%%%%%%%%%%%%
In the macroscopic time-space diagram, the platoon composed of vehicles $i_{\rm a}, i_{\rm a} + 1, i_{\rm a} + 2, \ldots$ run at the low velocity $v_{\rm a, mac}$ for an infinitely long time. To restrict secondary jams, we should ensure that this platoon retains linear string stability. Therefore, we should keep $v_{\rm a, mac}$ greater than or equal to $v_{\rm cr}$. That is,
\begin{align}
v_{\rm a, mac} = \dfrac{ (v_{\rm J} - v_{\rm R}) v_{\rm ini} + (v_{\rm R} - v_{\rm S}) v_{\rm J} }{ v_{\rm J} - v_{\rm S} } \ge v_{\rm cr}.
\label{eq:cond_no_secondary_jam}
\end{align}
In the case of the IDM ($v_{\rm J}=0$), this condition can be simplified as follows:
\begin{align}
v_{\rm a, mac} = \dfrac{ v_{\rm R} }{ v_{\rm S} } v_{\rm ini} \ge v_{\rm cr}.
\label{eq:cond_no_secondary_jam_simplified}
\end{align}
Condition~(\ref{eq:cond_no_secondary_jam}) or (\ref{eq:cond_no_secondary_jam_simplified}) can be used as the linear string stability condition for JAD.
In the case of a constant wide moving jam ($v_{\rm R} = v_{\rm S}$), condition~(\ref{eq:cond_no_secondary_jam}) becomes $v_{\rm ini} \ge v_{\rm cr}$ regardless of the value of $v_{\rm J}$.

Here we consider the case of finite systems. We add a horizontal line $v_{\rm cr} = 20.13\,{\rm m}/{\rm s}$ in Fig.~\ref{fig:vini_va_secondary_jam}.
This $v_{\rm cr}$ is greater than all the threshold values of $v_{\rm a}$ determining the presence/absence of secondary jams under $i_{\rm a} = 2 N / 5 + 1$ and $N \in \left\{10^3, 10^4, 10^5\right\}$.
This relationship between $v_{\rm cr}$ and the threshold values of $v_{\rm a}$ suggests that
\begin{align}
v_{\rm a} \ge v_{\rm cr}
\label{eq:cond_v_a_ge_v_cr}
\end{align}
is a suitable condition for suppressing secondary jams under finite systems.
Note the relationship between conditions~(\ref{eq:cond_v_a_ge_v_cr}) and (\ref{eq:cond_no_secondary_jam}).
Because $v_{\rm a}$ converges to $v_{\rm a, mac}$ as $i_{\rm a}$ goes to infinity, condition~(\ref{eq:cond_v_a_ge_v_cr}) converges to condition~(\ref{eq:cond_no_secondary_jam}) as both $i_{\rm a}$ and $N-i_{\rm a}$ go to infinity.
%%%%%%%%%%%%%%%%%%%%%%%%%%%%%%%%%%%%%%%%%%%%%%%%%%%%%%%%%%%%%%%%%%%%%%%%%%%%%%%%%%%%%%%%%%%%%%%%%%%%
\subsection{\label{subsec:influence_param_secondary_jams} Behaviors for a semi-infinite system}
%%%%%%%%%%%%%%%%%%%%%%%%%%%%%%%%%%%%%%%%%%%%%%%%%%%%%%%%%%%%%%%%%%%%%%%%%%%%%%%%%%%%%%%%%%%%%%%%%%%%
We investigate the influence of the parameters of the IDM ($a$, $b$, and $T$) and the initial velocity $v_{\rm ini}$ on the behavior in the semi-infinite system in which $i_{\rm a}$ and $N-i_{\rm a}$ are infinite.
We categorize the behavior into three cases:
\begin{description}
\item[Free (F)] Vehicle 1's perturbation does not grow into a wide moving jam. JAD is not necessary in this case.
\item[No secondary jam (NSJ)] Vehicle 1's perturbation grows into a wide moving jam. JAD removes it and restricts the occurrence of secondary jams.
\item[Secondary jam (SJ)] Vehicle 1's perturbation grows into a wide moving jam. JAD removes it, but causes secondary jams.
\end{description}
In judging the behavior, we utilize condition~(\ref{eq:cond_no_secondary_jam_simplified}).
Recall that it is not possible to perform numerical simulations in the semi-infinite system. Therefore, we also utilize the numerical simulations under a finite and sufficiently large $N$ to determine the behavior.
In the numerical simulations, we set $N=10^3$ and the maximal time of a run to $8 N\,{\rm s}$. Vehicle 1 causes the initial perturbation, but JAD is not activated throughout the run of the numerical simulations. Accordingly, if a wide moving jam arises as a consequence of vehicle 1's perturbation in a run, it will propagate in the upstream direction.

After completing a run, we judge the behavior of the semi-infinite system over two steps. In the first step, if no jam has arisen in the run, we judge the behavior to be F and skip the second step. Otherwise, we proceed to the second step whether NSJ or SJ occurred. Note that we judge a wide moving jam to have arisen if $v_N(t)$ falls below $1\,{\rm m}/{\rm s}$ at least once in the run.

In the second step, we judge the behavior to be NSJ if condition~(\ref{eq:cond_no_secondary_jam_simplified}) is satisfied. Otherwise, we judge that the behavior to be SJ.
In applying condition~(\ref{eq:cond_no_secondary_jam_simplified}), we need $v_{\rm S}$, $v_{\rm R}$, and $v_{\rm cr}$.
We obtain $v_{\rm cr}$ numerically using condition~(\ref{eq:string_stability_IDM}), and calculate $v_{\rm S}$ as the slope connecting the time-space points in which the two vehicles (vehicles $N-100$ and $N$) enter the wide moving jam in the run. We identify a vehicle's entry into the jam according to the threshold velocity $1\,{\rm m}/{\rm s}$. We also calculate $v_{\rm R}$ as the slope connecting the time-space points in which the two vehicles escape from the jam using the same threshold velocity.

\begin{figure}[t]
\centering
\includegraphics[width=\hsize]{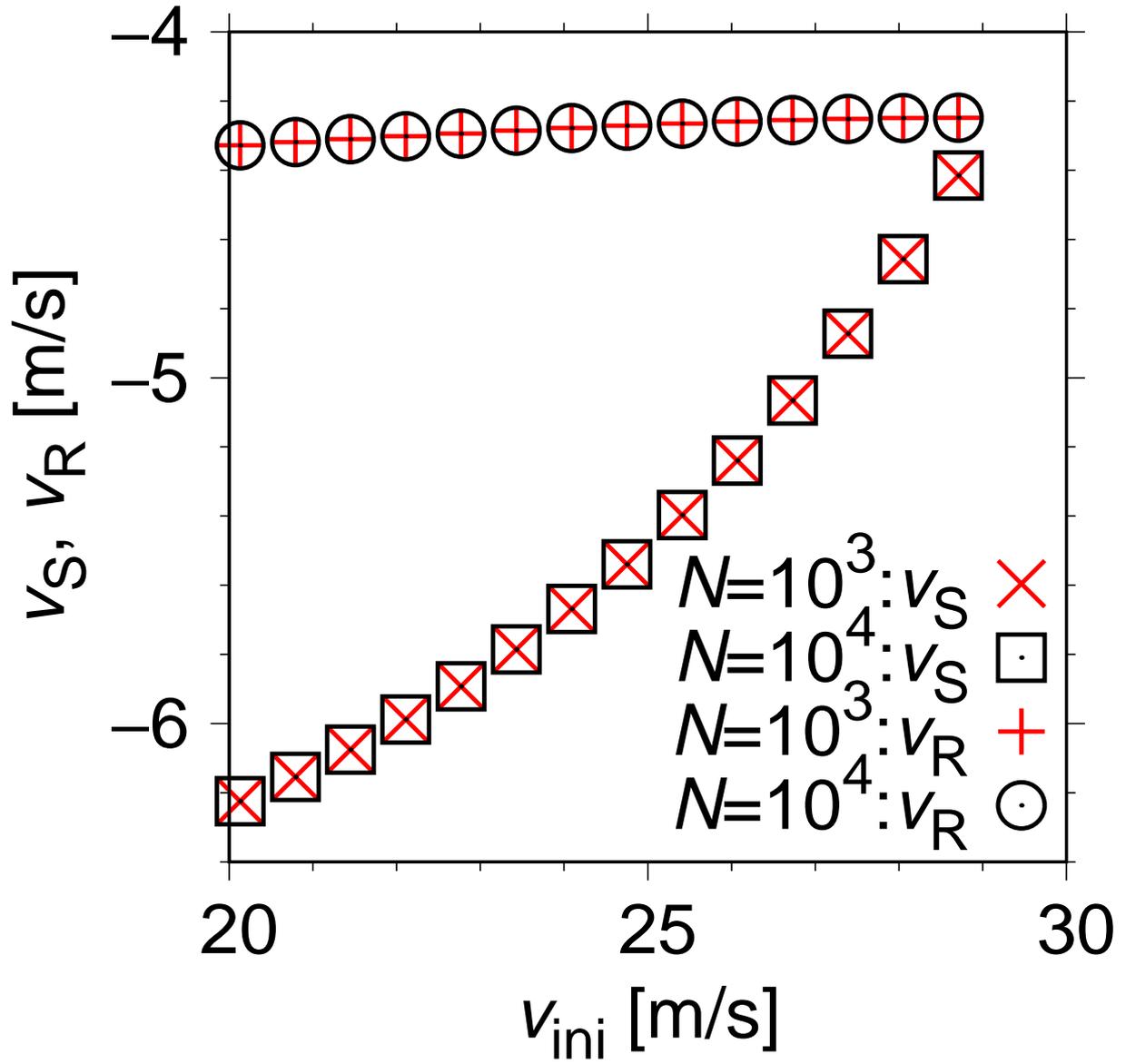}
\caption{
Velocity of the upstream tail of the jam $v_{\rm S}$ and the velocity of the downstream head of the jam $v_{\rm R}$ as functions of the initial velocity $v_{\rm ini}$ under $N\in\left\{10^3, 10^4\right\}$.
The range of $v_{\rm ini}$ is given by $v_{\rm ini} = v_{\rm cr} + j (v_0 - v_{\rm cr}) / 20$ ($j = 0, 1, \ldots, 19$), where $v_{\rm cr} = 20.13\,{\rm m}/{\rm s}$.
$v_{\rm S}$ and $v_{\rm R}$ are obtained for $0 \le j \le 13$ and are not obtained for $14 \le j \le 19$.
Only those values of $v_{\rm S}$ and $v_{\rm R}$ obtained by the numerical simulations are shown.
}
\label{fig:v_S_v_R}
\end{figure}

We now check the validity of using a finite system of $N=10^3$ in judging the behavior under a semi-infinite system.
We compare $v_{\rm S}$ and $v_{\rm R}$ as functions of $v_{\rm ini}$ obtained from the numerical simulations under $N=10^3$ with those obtained under $N=10^4$.
We set $v_{\rm cr} = 20.13$ ${\rm m}/{\rm s}$ and $v_{\rm ini} = v_{\rm cr} + j(v_0 - v_{\rm cr})/20$ ($j = 0, 1, \ldots, 19$).
The comparison results are presented in Fig.~\ref{fig:v_S_v_R}.
Note that we only plot $v_{\rm S}$ and $v_{\rm R}$ only when they have been obtained.
In Fig.~\ref{fig:v_S_v_R}, $v_{\rm S}$ and $v_{\rm R}$ are obtained for $0 \le j \le 13$ and are not obtained for $14 \le j \le 19$. Hence, the threshold value of $v_{\rm ini}$ that determines whether or not to obtain $v_{\rm S}$ and $v_{\rm R}$ under $N=10^3$ coincides with that under $N=10^4$. This coincidence suggests that $N=10^3$ is sufficiently large for dividing the behavior into F or NSJ/SJ.
Moreover, $v_{\rm S}$ and $v_{\rm R}$ for the case $N=10^3$ agree with those for $N=10^4$, which suggests that $v_{\rm S}$ and $v_{\rm R}$ are constant when $N=10^3$ or above.
$v_{\rm S}$ and $v_{\rm R}$ under $N=\infty$ are therefore given by their values in the case $N=10^3$, as mentioned above, and $v_{\rm cr}$ is independent of $N$. Accordingly, we can check condition~(\ref{eq:cond_no_secondary_jam_simplified}) under $N=\infty$ by setting $N=10^3$.
Thus, we conclude that using a finite system of $N=10^3$ is valid in terms of judging the behavior of a semi-infinite system.

The influence of $a$, $b$, $T$, and $v_{\rm ini}$ on $v_{\rm cr}$, $v_{\rm R}$, $v_{\rm S}$, and $v_{\rm R}/v_{\rm S}$ was investigated as follows.
We set the ranges of $a$ to $0.5 \le a \le 1.5\,{\rm m}/{\rm s}^2$. For each value of $a$, we numerically calculate $v_{\rm cr}$ according to condition~(\ref{eq:string_stability_IDM}).
Then, we set $v_{\rm ini} = v_{\rm min} + j(v_0 - v_{\rm min})/20$ ($j = 0, 1, \ldots, 19$), where $v_{\rm min} = 10\,{\rm m}/{\rm s}$, and fix those parameters not explicitly mentioned to the same values as in Table~\ref{tab:param}.
We obtain $v_{\rm S}$, $v_{\rm R}$, and $v_{\rm R}/v_{\rm S}$ through numerical simulations with $N=10^3$. In this way, we obtain $v_{\rm cr}$ as a function of $a$ and obtain $v_{\rm S}$, $v_{\rm R}$, and $v_{\rm R}/v_{\rm S}$ as functions of $v_{\rm ini}$ and $a$.
In the same way, we set $1 \le b \le 2\,{\rm m}/{\rm s}^2$ and obtain $v_{\rm cr}$ as a function of $b$ and $v_{\rm S}$, $v_{\rm R}$, and $v_{\rm R}/v_{\rm S}$ as functions of $v_{\rm ini}$ and $b$.
Additionally, we set $0.5 \le T \le 1.5\,{\rm s}$ and obtain $v_{\rm cr}$ as a function of $T$ and $v_{\rm S}$, $v_{\rm R}$, and $v_{\rm R}/v_{\rm S}$ as functions of $v_{\rm ini}$ and $T$.

\begin{figure}[t]
\centering
\includegraphics[width=\hsize]{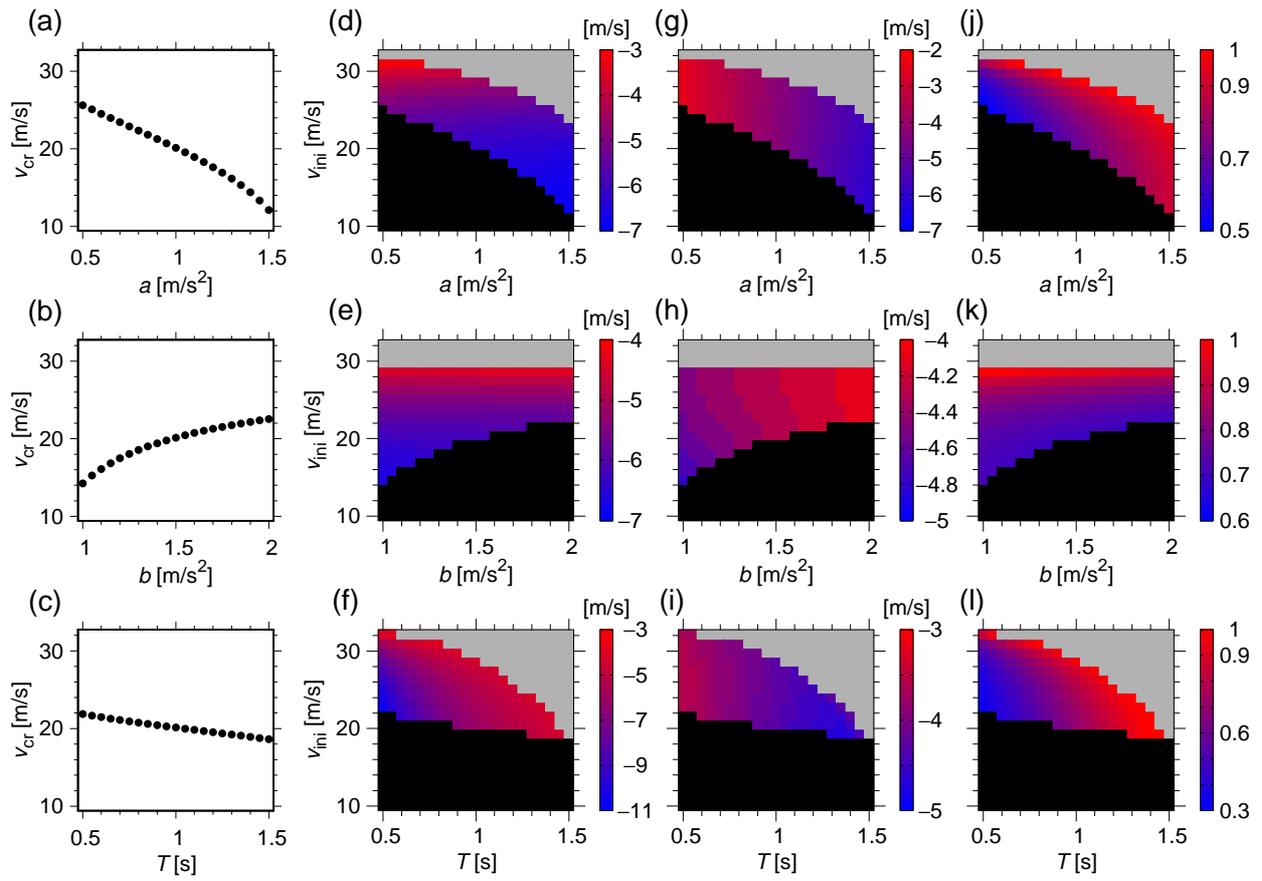}
\caption{
(a)--(c): $v_{\rm cr}$ as a function of (a) $a$, (b) $b$ and (c) $T$.
(d)--(f): $v_{\rm S}$ as a function of $v_{\rm ini}$ and (d) $a$, (e) $b$ and (f) $T$.
(g)--(i): $v_{\rm R}$ as a function of $v_{\rm ini}$ and (g) $a$, (h) $b$ and (i) $T$.
(j)--(l): $v_{\rm R}/v_{\rm S}$ as a function of $v_{\rm ini}$ and (j) $a$, (k) $b$ and (l) $T$.
(d)--(l): Gray regions represent the regions in which the wide moving jam is absent. Black regions represent the regions in which $v_{\rm ini} < v_{\rm cr}$.
}
\label{fig:v_R_v_S_v_cr}
\end{figure}

Figure~\ref{fig:v_R_v_S_v_cr} shows the results.
The regions of behavior F (that is, the regions in which the wide moving jam is absent) are shown in gray in Figs.~\ref{fig:v_R_v_S_v_cr}(d)--(l).
The regions in which $v_{\rm ini} < v_{\rm cr}$ are shown in black in Figs.~\ref{fig:v_R_v_S_v_cr}(d)--(l).
$v_{\rm cr}$ decreases with respect to $a$ and $T$ and increases with respect to $b$, as shown in Figs.~\ref{fig:v_R_v_S_v_cr}(a)--(c). The influence of $a$ on $v_{\rm cr}$ is greater than that on $T$.
$v_{\rm S}$ is approximately constant with respect to $a$ and $b$, and increases with respect to $T$ and $v_{\rm ini}$, as shown in Figs.~\ref{fig:v_R_v_S_v_cr}(d)--(f).
$v_{\rm R}$ decreases with respect to $a$ and $T$, increases with respect to $b$, and is approximately constant with respect to $v_{\rm ini}$, as shown in Figs.~\ref{fig:v_R_v_S_v_cr}(g)--(i).
$v_{\rm R}/v_{\rm S}$ increases with respect to $a$, $T$, and $v_{\rm ini}$ and is approximately constant with respect to $b$, as shown in Figs.~\ref{fig:v_R_v_S_v_cr}(j)--(l).

\begin{figure}[t]
\centering
\includegraphics[width=\hsize]{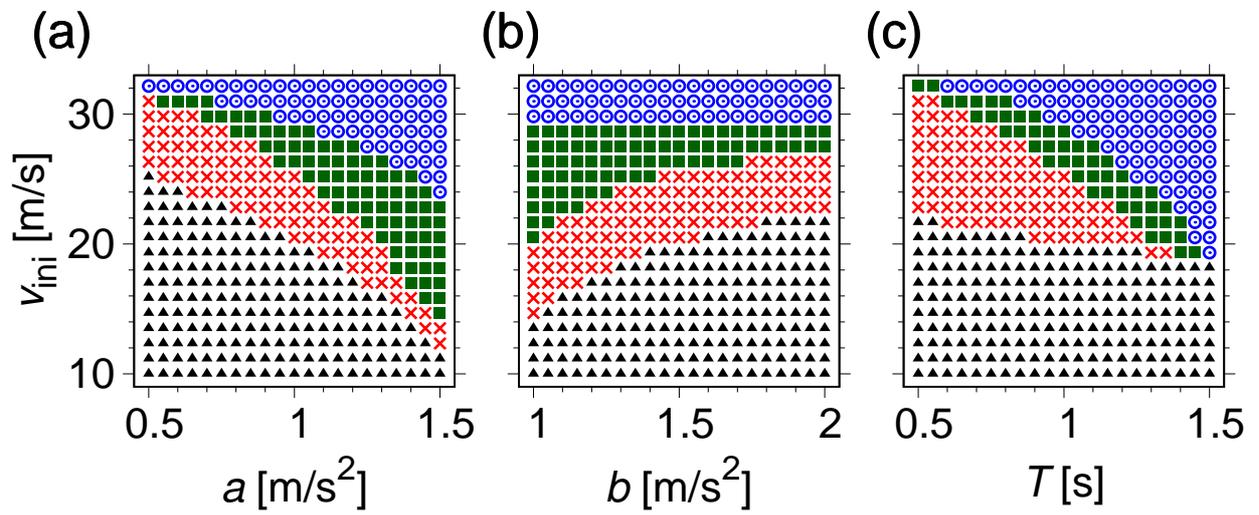}
\caption{
Behavior of the semi-infinite system in which $i_{\rm a}$ and $N-i_{\rm a}$ are infinite as a function of $v_{\rm ini}$ and (a) $a$, (b) $b$ and (c) $T$.
The other IDM parameters are fixed to the values in Table~\ref{tab:param}.
Behaviors F, NSJ and SJ are depicted by blue open circles, green filled squares and red crosses, respectively.
The regions in which $v_{\rm ini} < v_{\rm cr}$ are depicted by black filled triangles.
}
\label{fig:consequences}
\end{figure}

Using the results shown in Fig.~\ref{fig:v_R_v_S_v_cr} (that is, the absence or occurrence of the wide moving jam and the values of $v_{\rm cr}$, $v_{\rm R}$, and $v_{\rm S}$ under $N=10^3$), we ascertain the influence of $a$, $b$, $T$, and $v_{\rm ini}$ on the behavior when $i_{\rm a}$ and $N - i_{\rm a}$ are infinite, as shown in Fig.~\ref{fig:consequences}. We depict F, NSJ, and SJ by blue open circles, green filled squares, and red crosses, respectively. Note that we depict the regions in which $v_{\rm ini} < v_{\rm cr}$ by black filled triangles.
All three behaviors F, NSJ, and SJ appear over wide ranges of $a$, $b$, and $T$. The existence of NSJ means that the semi-infinite system can recover from a wide moving jam through only a single vehicle's maneuvers. In particular, when the parameters of the IDM are set to the typical values for highway traffic ($a=1\,{\rm m}/{\rm s}^2$, $b=1.5\,{\rm m}/{\rm s}^2$, and $T=\,1{\rm s}$, as listed in Table~\ref{tab:param}), $v_{\rm ini}$ values that result in NSJ indeed exist, which supports the applicability of JAD without causing secondary jams.
We now highlight several details of Fig.~\ref{fig:consequences}. As $v_{\rm ini}$ increases from $v_{\rm cr}$ toward $v_0$, the behavior tends to be SJ, then NSJ, and finally F.
The region of NSJ becomes wider as $a$ increases, as shown in Fig.~\ref{fig:consequences}(a). We believe this tendency to be caused by $v_{\rm cr}$ decreasing monotonically against $a$ (Fig.~\ref{fig:v_R_v_S_v_cr}(a)) and $v_{\rm R}/v_{\rm S}$ increasing monotonically against $a$ and $v_{\rm ini}$ (Fig.~\ref{fig:v_R_v_S_v_cr}(j)).
The NSJ region becomes narrower as $b$ increases, as shown in Fig.~\ref{fig:consequences}(b). This dependence of NSJ on $b$ is likely to be caused by $v_{\rm cr}$ increasing monotonically against $b$ (Fig.~\ref{fig:v_R_v_S_v_cr}(b)), and $v_{\rm R}/v_{\rm S}$ approximately constant against $b$ (Fig.~\ref{fig:v_R_v_S_v_cr}(k)).
The value of $v_{\rm ini}$ that produces NSJ moves from nearly $v_0$ to nearly $v_{\rm cr}$ as $T$ increases, as shown in Fig.~\ref{fig:consequences}(c). We believe this dependence of NSJ on $T$ is caused by $v_{\rm R}/v_{\rm S}$ increasing monotonically against $T$ and $v_{\rm ini}$ (Fig.~\ref{fig:v_R_v_S_v_cr}(l)), and $v_{\rm cr}$ decreasing slightly against $T$ (Fig.~\ref{fig:v_R_v_S_v_cr}(c)).
%%%%%%%%%%%%%%%%%%%%%%%%%%%%%%%%%%%%%%%%%%%%%%%%%%%%%%%%%%%%%%%%%%%%%%%%%%%%%%%%%%%%%%%%%%%%%%%%%
\subsection{\label{subsec:other_semi_infinite_systems} Behavior in other semi-infinite systems}
%%%%%%%%%%%%%%%%%%%%%%%%%%%%%%%%%%%%%%%%%%%%%%%%%%%%%%%%%%%%%%%%%%%%%%%%%%%%%%%%%%%%%%%%%%%%%%%%%
Although we have mainly focused on JAD on a single-lane system without bottlenecks, many real highways have multiple lanes and bottlenecks.
We now investigate the behavior of semi-infinite systems with inflows from other lanes or a bottleneck by using or extending conditions~(\ref{eq:cond_no_secondary_jam}) and (\ref{eq:cond_no_secondary_jam_simplified}).
%%%%%%%%%%%%%%%%%%%%%%%%%%%%%%%%%%%%%%%%%%%%%%%%%%%%%%%%%%%%%%%%%%%%%%%%%%%%%%%%%%%%%%%%%%%%%%
\subsubsection{\label{subsubsec:multiple_lane_system} System with inflows from other lanes}
%%%%%%%%%%%%%%%%%%%%%%%%%%%%%%%%%%%%%%%%%%%%%%%%%%%%%%%%%%%%%%%%%%%%%%%%%%%%%%%%%%%%%%%%%%%%%%
We consider a multiple-lane system in which a single absorbing vehicle removes a wide moving jam propagating on one lane. When the absorbing vehicle produces a vacant space on this lane, vehicles in the neighboring lanes may enter this vacant space. These inflows extend the downstream head of the targeted traffic jam to the upstream direction. Therefore, the absorbing velocity in this multiple-lane system is smaller than that in single-lane systems.
As the traffic flows on the other lanes, we consider only the inflows to the vacant space. That is, we omit the time evolution of the traffic flows on the other lanes. This simplification does not adversely affect the investigation of the influence of inflows on the performance of JAD.

\begin{figure}[t]
\centering
\includegraphics[width=\hsize]{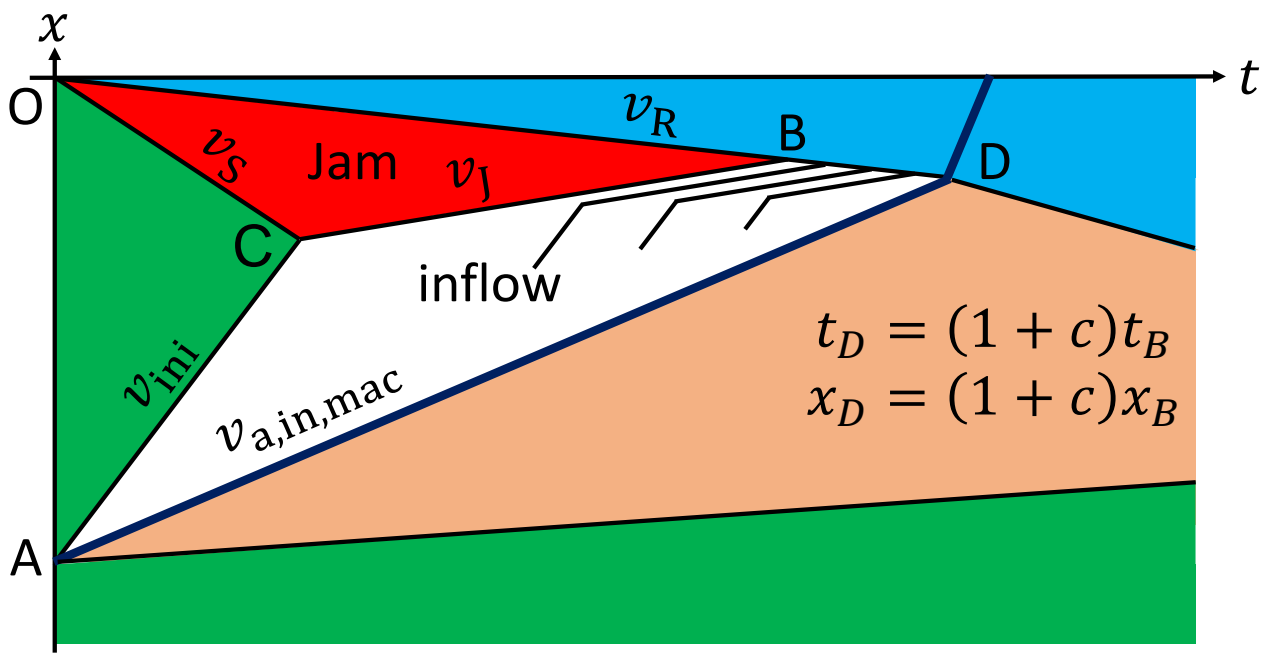}
\caption{Macroscopic view of JAD when vehicles in neighboring lanes enter the vacant space produced by the absorbing vehicle on a time-space diagram. These vehicles move the disappearing point of the wide moving jam from point B to D, where $t_{\rm D} = (1+c) t_{\rm B}$ and $x_{\rm D} = (1+c) x_{\rm B}$. $c$ denotes the degree of inflows. The absorbing vehicle goes from point A to D at velocity $v_{\rm a, in, mac}$ during the slow-in phase. Its velocity increases at point D, signifying the fast-out phase.}
\label{fig:macro_view_inflow}
\end{figure}

We depict a macroscopic view of JAD in this multiple-lane system, as shown in Fig.~\ref{fig:macro_view_inflow}. In this figure, the downstream head of a wide moving jam disappears not at point B but at point D because of the inflows. The absorbing vehicle goes from point A to D at the macroscopic absorbing velocity $v_{\rm a, in, mac}$. Note that we assume that the absorbing vehicle can estimate the time and position of point D from the initial time.
We set the time and position of D to:
\begin{align}
t_{\rm D} = (1+c) t_{\rm B}, \ \ \ \ x_{\rm D} = (1+c) x_{\rm B}.
\label{eq:t_D_x_D}
\end{align}
The parameter $c$ denotes the ratio of the inflowing vehicles to the vehicles which originally enter the jam in the no-inflow case (that is, vehicles $1, 2, \ldots, i_{\rm a} - 1$). Hence, the number of inflowing vehicles is approximately given by $c i_{\rm a}$. If $c=0$, the system is identical to the single-lane system.

$v_{\rm a, in, mac}$ is given as follows by referring to Fig.~\ref{fig:macro_view_inflow}:
\begin{align}
v_{\rm a, in, mac} = \dfrac{v_{\rm a, mac} + c v_{\rm R}}{1+c}.
\label{eq:v_a_in_mac}
\end{align}
Because $v_{\rm a, in, mac}$ is equal to or greater than zero, $c$ should satisfy the following condition:
\begin{align}
0 \le c \le - \dfrac{v_{\rm a, mac}}{v_{\rm R}}.
\label{eq:range_c}
\end{align}
When $v_{\rm J} = 0$, $v_{\rm a, in, mac}$ and the condition of $c$ can be simplified as:
\begin{align}
v_{\rm a, in, mac} = \dfrac{v_{\rm R}}{1+c} \left( \dfrac{v_{\rm ini}}{v_{\rm S}} + c \right),
\label{eq:v_a_in_mac_simplified}
\end{align}
\begin{align}
0 \le c \le - \dfrac{v_{\rm ini}}{v_{\rm S}}.
\label{eq:range_c_simplified}
\end{align}
The platoon composed of the absorbing vehicle and vehicles upstream of it is linearly string stable if
\begin{align}
v_{\rm a, in, mac} \ge v_{\rm cr}.
\label{eq:v_a_in_mac_stability_cond}
\end{align}

We investigate the influence of $v_{\rm ini}$ and $c$ on the behaviors of this multiple-lane system. We reuse F, NSJ, and SJ for a single-lane system defined in Sec.~\ref{subsec:influence_param_secondary_jams}.
We determine the behavior in two steps. If no traffic jam occurs in the single-lane system of system size $10^3$ which has been treated in Sec.~\ref{subsec:influence_param_secondary_jams}, we judge the behavior to be F and skip the second step. In the second step, if condition~(\ref{eq:v_a_in_mac_stability_cond}) is satisfied, we judge the behavior to be NSJ. Otherwise, we judge the behavior to be SJ.

\begin{figure}[t]
\centering
\includegraphics[width=\hsize]{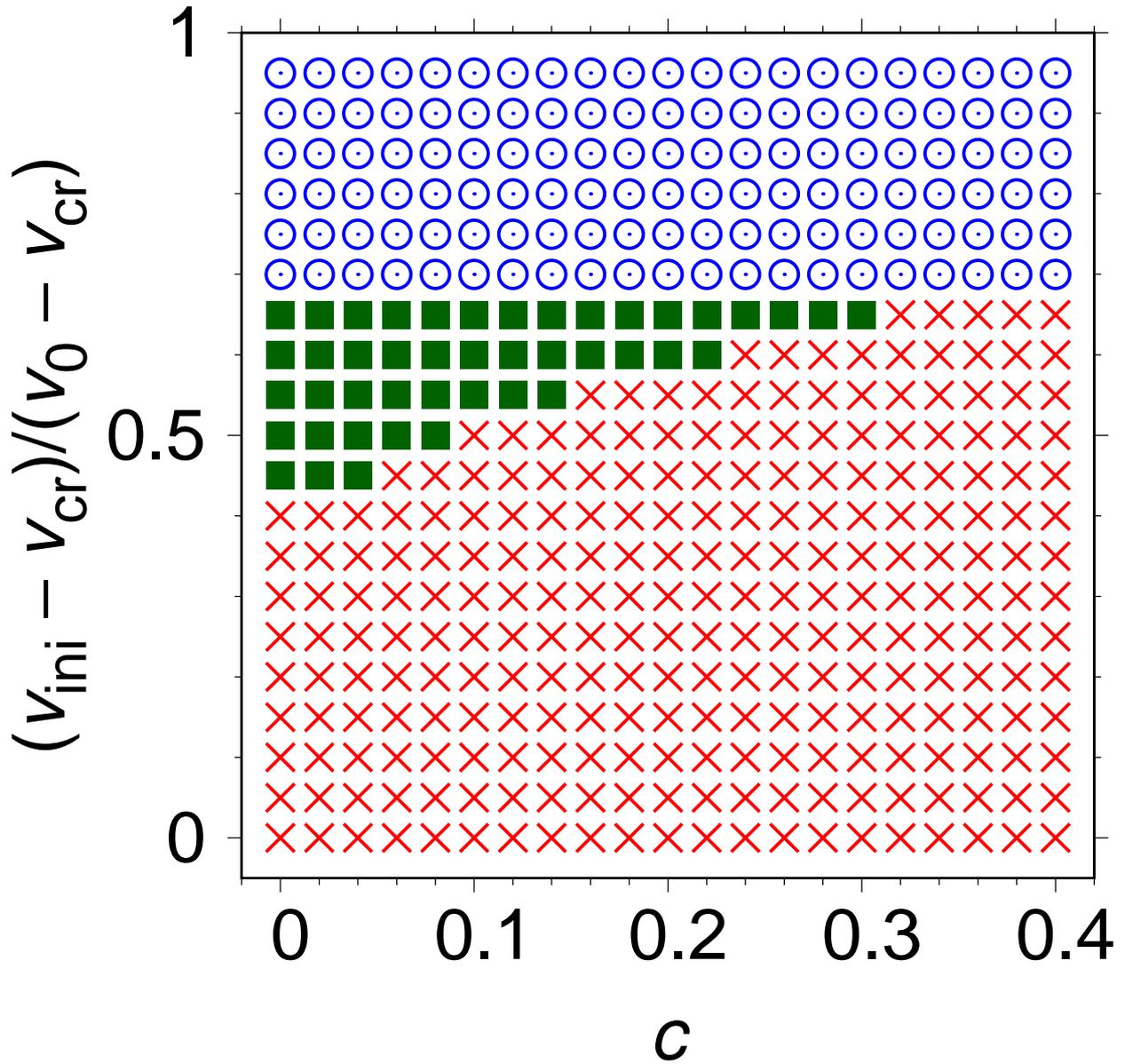}
\caption{
Behavior of a semi-infinite system with inflows of vehicles from neighboring lanes to the vacant space produced by the absorbing vehicle, as a function of $v_{\rm ini}$ and $c$.
The IDM parameters are fixed to the values in Table~\ref{tab:param}.
$v_{\rm cr}$ is set to $20.13\,{\rm m}/{\rm s}$.
The vertical axis is normalized according to $(v_{\rm ini} - v_{\rm cr}) / (v_0 - v_{\rm cr})$.
Behaviors F, NSJ and SJ are depicted by blue open circles, green filled squares and red crosses, respectively.
}
\label{fig:behavior_inflow}
\end{figure}

Figure~\ref{fig:behavior_inflow} shows the behavior of the multiple-lane system as a function of $v_{\rm ini}$ and $c$.
In this figure, we set the IDM parameters as listed in Table~\ref{tab:param} and set $v_{\rm cr} = 20.13\,{\rm m}/{\rm s}$.
We set the range of $v_{\rm ini}$ to $v_{\rm cr} \le v_{\rm ini} < v_0$, and the range of $c$ to $0 \le c \le 0.4$.
We used the same symbols as those in Fig.~\ref{fig:consequences} for representing the behaviors.
We normalized the vertical axis of this figure according to $(v_{\rm ini} - v_{\rm cr}) / (v_0 - v_{\rm cr})$.
As $c$ increases, the minimum value of $v_{\rm ini}$ producing NSJ increases, and the range of $v_{\rm ini}$ producing NSJ becomes narrower. Nevertheless, NSJ remains under $c \le 0.3$. In addition, when $c \le 0.14$, the range of $v_{\rm ini}$ producing NSJ maintains three-fifths of that under no inflow case ($c = 0$).
Note that condition~(\ref{eq:range_c_simplified}) was always satisfied in NSJ and SJ regions under the aforementioned parameter settings.
%%%%%%%%%%%%%%%%%%%%%%%%%%%%%%%%%%%%%%%%%%%%%%%%%%%%%%%%%%%%%%%%%%%%%%%%%%%%%
\subsubsection{\label{subsubsec:system_bottleneck} System with a bottleneck}
\begin{figure}[t]
\centering
\includegraphics[width=\hsize]{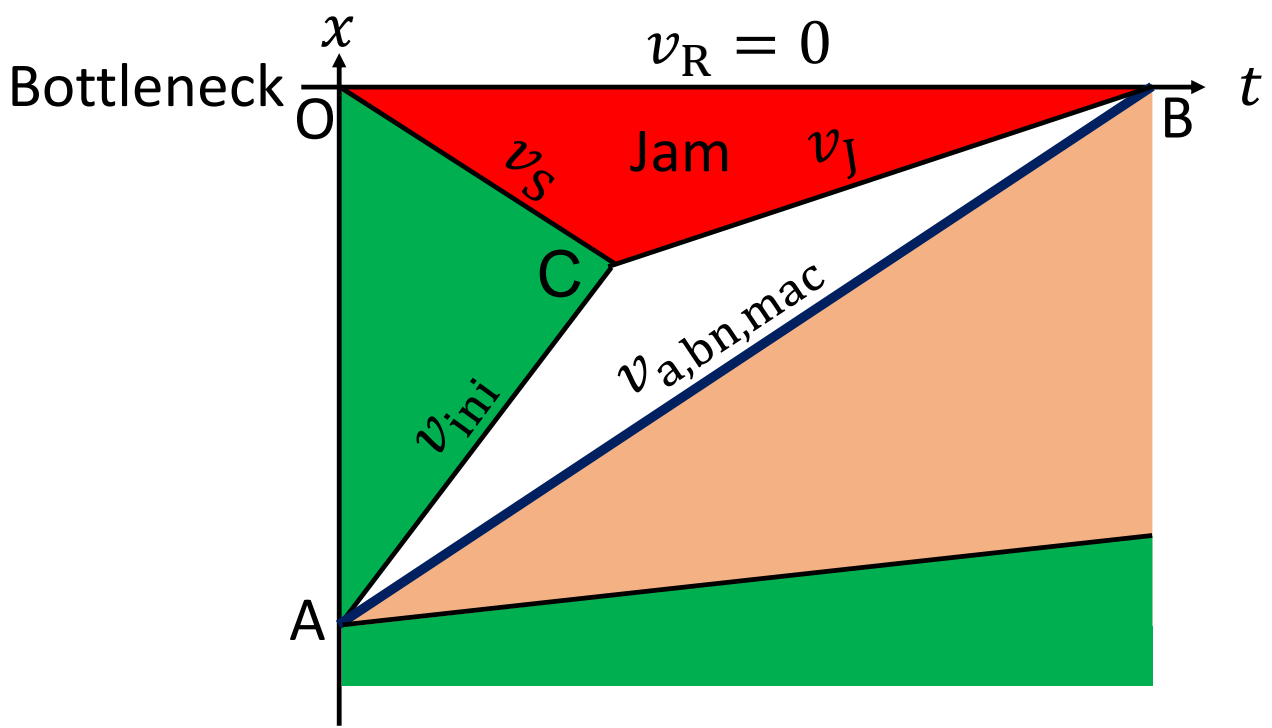}
\caption{Macroscopic view of JAD for removing a traffic jam fixed at a bottleneck on a time-space diagram. The bottleneck is placed at $x=0$. We represent this jam by the triangle OBC which has $v_{\rm R} = 0$. The absorbing vehicle goes from point A to B at velocity $v_{\rm a, bn, mac}$ during the slow-in phase. Its velocity increases at point B, signifying the fast-out phase.}
\label{fig:macro_view_bottleneck}
\end{figure}

%%%%%%%%%%%%%%%%%%%%%%%%%%%%%%%%%%%%%%%%%%%%%%%%%%%%%%%%%%%%%%%%%%%%%%%%%%%%%
Various scenarios of the maneuvers of one or more CVs were proposed and analyzed for mitigating traffic jams fixed at a bottleneck~\citep{Han2017}. Based on the basic scenario of Ref.~\citep{Han2017}, we focus on a single-lane system with a bottleneck, as shown in the macroscopic view on a time-space diagram (Fig.~\ref{fig:macro_view_bottleneck}). A bottleneck is placed at $x=0$, which is a sag, not an on-ramp. Therefore, there is no inflow from other roads at this bottleneck. Contrary to the scenarios of wide moving jams, vehicle 1 does not produce its initial perturbation. Instead of vehicle 1, this bottleneck triggers off traffic breakdown (a drop of flow rates) and causes a traffic jam~\citep{Chen2014,Han2017}. The jam arises from O $(0, 0)$ and its downstream head is fixed at this bottleneck (that is, $v_{\rm R} = 0$).
Before vehicles enter the jam, they run at velocity $v_{\rm ini}$.
When they enter it, their velocity becomes $v_{\rm J}$.
We assume that the jam slows down vehicles:
\begin{align}
v_{\rm J} < v_{\rm ini}.
\label{eq:v_J_v_ini_cond}
\end{align}
We also assume that traffic flows inside the jam are in equilibrium with velocity $v_{\rm J}$ and equilibrium density $\rho_{\rm e}(v_{\rm J})$, where $\rho_{\rm e}(v)$ is given by:
\begin{align}
\rho_{\rm e}(v) = \dfrac{1}{s_{\rm e}(v) + d}.
\label{eq:rho_e_v}
\end{align}
After vehicles escape from the jam, their velocity becomes higher.
$v_{\rm S}$ is given by:
\begin{align}
v_{\rm S} = \dfrac{ \rho_{\rm e}(v_{\rm ini}) v_{\rm ini} - \rho_{\rm e}(v_{\rm J}) v_{\rm J} }{ \rho_{\rm e}(v_{\rm ini}) - \rho_{\rm e}(v_{\rm J})}.
\label{eq:v_S_bn}
\end{align}

The absorbing vehicle goes from point A to B at the macroscopic absorbing velocity $v_{\rm a, bn, mac}$ for dissolving the jam~\citep{Han2017}. We obtain $v_{\rm a, bn, mac}$ by substituting $v_{\rm R} = 0$ for condition~(\ref{eq:cond_no_secondary_jam}):
\begin{align}
v_{\rm a, bn, mac} = \dfrac{v_{\rm J} (v_{\rm ini} - v_{\rm S}) }{ v_{\rm J} - v_{\rm S} }.
\label{eq:v_a_bn_mac}
\end{align}
The linear string stability condition of the platoon composed of the absorbing vehicle and the vehicles upstream of it is given by:
\begin{align}
v_{\rm a, bn, mac} \ge v_{\rm cr}.
\label{eq:v_a_bn_mac_stability_cond}
\end{align}
As secondary jams, we only consider those caused by the instabilities in running at the absorbing velocity $v_{\rm a, bn, mac}$. We do not consider the next traffic jams arising at $x=0$, which Ref.~\citep{Han2017} analyzed in detail.

In investigating the behavior of the system, we consider only the situations in which traffic breakdown occurs.
Therefore, the flow rate inside the jam should be lower than the flow rate upstream of the jam:
\begin{align}
\rho_{\rm e}(v_{\rm J})v_{\rm J} < \rho_{\rm e}(v_{\rm ini})v_{\rm ini}.
\label{eq:traffic_breakdown_cond}
\end{align}
Because condition~(\ref{eq:v_J_v_ini_cond}) is satisfied, $\rho_{\rm e}(v_{\rm ini})$ is smaller than $\rho_{\rm e}(v_{\rm J})$ in the IDM. Therefore, condition~(\ref{eq:traffic_breakdown_cond}) is rewritten as:
\begin{align}
v_{\rm S} < 0.
\label{eq:capacity_drop_cond_v_S}
\end{align}
Accordingly, we investigate the behavior of the system under conditions~(\ref{eq:v_J_v_ini_cond}) and~(\ref{eq:capacity_drop_cond_v_S}).
We categorize the behavior into the two cases.
\begin{description}
\item[No secondary jam (NSJ)] JAD removes the traffic jam and restricts secondary jams.
\item[Secondary jam (SJ)] JAD removes the traffic jam, but causes secondary jams.
\end{description}
We judge the behavior to be NSJ if condition~(\ref{eq:v_a_bn_mac_stability_cond}) is satisfied. Otherwise, we judge the behavior to be SJ.

\begin{figure}[t]
\centering
\includegraphics[width=\hsize]{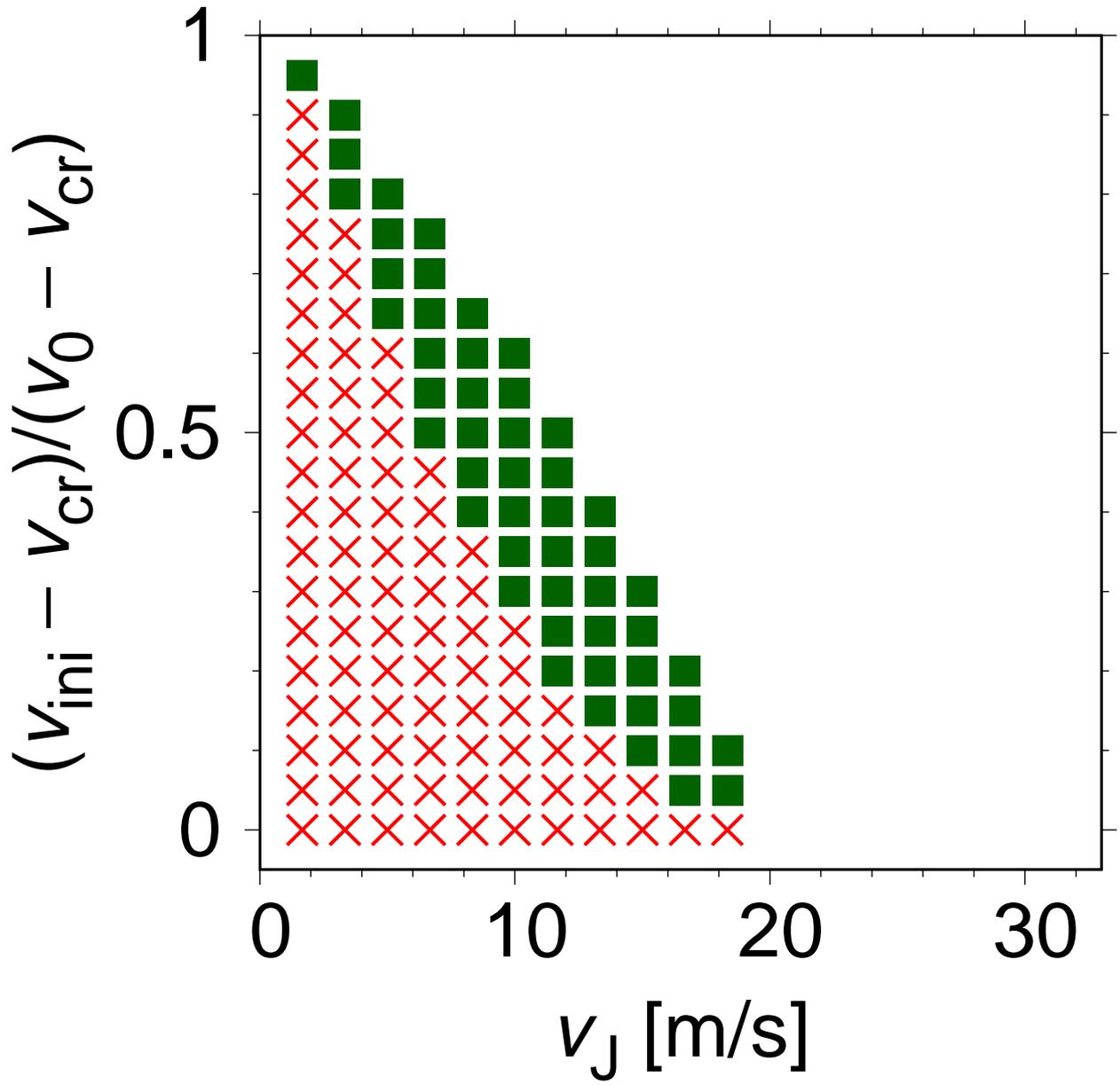}
\caption{
Behavior of a semi-infinite system containing a bottleneck as a function of $v_{\rm ini}$ and $v_{\rm J}$.
The IDM parameters are fixed to the values in Table~\ref{tab:param}.
$v_{\rm cr}$ is set to $20.13\,{\rm m}/{\rm s}$.
The vertical axis is normalized according to $(v_{\rm ini} - v_{\rm cr}) / (v_0 - v_{\rm cr})$.
Behaviors NSJ and SJ are depicted by green filled squares and red crosses, respectively.
Note that we only depict the points where both conditions~(\ref{eq:v_J_v_ini_cond}) and~(\ref{eq:capacity_drop_cond_v_S}) are satisfied.
}
\label{fig:behavior_bottleneck}
\end{figure}

We investigate the dependence of the behavior of the system on $v_{\rm ini}$ and $v_{\rm J}$.
We set the ranges of the parameters to $v_{\rm cr} \le v_{\rm ini} < v_0$ and $0 < v_{\rm J} < v_0$. The IDM parameters are listed in Table~\ref{tab:param}, and $v_{\rm cr}$ is set to $20.13\,{\rm m}/{\rm s}$.
We show the result in Fig.~\ref{fig:behavior_bottleneck}. The vertical axis of this figure is normalized according to $(v_{\rm ini} - v_{\rm cr}) / (v_0 - v_{\rm cr})$. We depict NSJ and SJ by green filled squares and red crosses, respectively. Note that we only depict the results in which both conditions~(\ref{eq:v_J_v_ini_cond}) and~(\ref{eq:capacity_drop_cond_v_S}) are satisfied.
Figure~\ref{fig:behavior_bottleneck} shows that both NSJ and SJ exist in wide ranges of $v_{\rm ini}$ and $v_{\rm J}$.
As $v_{\rm ini}$ increases, the behavior tends to be from SJ to NSJ.
As $v_{\rm J}$ increases, the behavior also tends to be from SJ to NSJ.
The existence of NSJ denotes that JAD can remove a traffic jam fixed at a bottleneck and restrict secondary jams.
%%%%%%%%%%%%%%%%%%%%%%%%%%%%%%%%%%%%%%%%%%%%%%%%%%%%%%%%%%%%%%%%%%%%%
\section{\label{sec:discussion}Discussion}
%%%%%%%%%%%%%%%%%%%%%%%%%%%%%%%%%%%%%%%%%%%%%%%%%%%%%%%%%%%%%%%%%%%%%
We set a single-lane road of infinite length without any loops or bottlenecks, and set all vehicles except for the absorbing vehicle to be HDVs. In this system, we have constructed conditions~(\ref{eq:cond_no_secondary_jam}) and (\ref{eq:cond_no_secondary_jam_simplified}) for removing a wide moving jam by JAD and suppressing the occurrence of secondary jams.
To construct these conditions, we applied the linear string stability condition~\citep{Wilson2008b, Treiber2013} to the macroscopic spatiotemporal structure of JAD.
Additionally, we have numerically confirmed that condition~(\ref{eq:cond_v_a_ge_v_cr}), which relates to finite systems, restricts secondary jams in finite systems composed of $10^3$--$10^5$ vehicles.
We have categorized the behavior of the semi-infinite system into the three cases: Free (F, no wide moving jam occurs), No secondary jam (NSJ, a wide moving jam occurs and JAD removes it without causing secondary jams), and Secondary jam (SJ, a wide moving jam occurs and JAD removes it, but causes secondary jams).
Utilizing condition~(\ref{eq:cond_no_secondary_jam_simplified}) and performing numerical simulations without activating JAD under $N=10^3$, we found that F, NSJ, and SJ exist widely on $a$--$v_{\rm ini}$ plane, $b$--$v_{\rm ini}$ plane, and $T$--$v_{\rm ini}$ plane.

The existence of NSJ regions guarantees that a single vehicle is able to return the traffic flow from a wide moving jam to a free flow in the semi-infinite system.
The existence of NSJ regions under typical IDM parameter values suggests that JAD is applicable under a suitable initial velocity (or initial density) of the traffic flow.

We have also constructed conditions for restricting the occurrence of secondary jams in other semi-infinite systems: the system with inflows from other lanes and the system with a bottleneck.
We have demonstrated that NSJ regions also exist in these other systems. The existence of NSJ regions shows that JAD is robust against the occurrence of secondary jams in these more complex systems.

Conditions~(\ref{eq:cond_no_secondary_jam}) and (\ref{eq:cond_no_secondary_jam_simplified}) may be applicable to other methods for easing traffic jams using car-following models, such as SPECIALIST with a car-following model~\citep{Wang2016JITS}, eco-driving~\citep{Chen2014TRR, He2015TRC}, and other similar driving methods~\citep{Wu2017, Ghiasi2017}.

Finally, we mention some ideas for potential future work.
The conditions developed in this paper is those for semi-infinite systems. Providing more accurate theoretical supports for restricting secondary jams in finite systems is challenging and warrants further studies.

We have assumed that the absorbing vehicle already knows the spatiotemporal evolution of the targeted traffic jam from the initial time. Predicting traffic jams costs a certain estimating time~\citep{Hegyi2013}, which may restrict the performance of JAD. Incorporating jam predictions into JAD will further clarify the robustness of JAD and warrants future work.

Although a way to select the absorbing vehicle was proposed~\citep{He2017}, we have designated the absorbing vehicle from the initial time in this paper. Selecting the most appropriate vehicle to perform JAD according to traffic conditions will improve the performance to dissipate traffic jams and stabilize the traffic flows upstream of it, and is worthy of further studies.

In treating a traffic jam whose downstream head is fixed at a bottleneck, we have imposed a considerably strong assumption that a traffic flow inside the jam is in equilibrium. Developing JAD for removing traffic jams out of equilibrium, for instance, traffic jams with fluctuation of density and velocity inside them, will improve the robustness of JAD, and warrants further studies.

In treating a multiple-lane system, we have only taken into account the inflows from other lanes. Constructing the theories of JAD for entire multiple-lane systems by considering traffic jams propagating on multiple lanes, lane-change rules, and heterogeneous traffic states among lanes will contribute to further understanding of the robustness of JAD.

We have set the traffic flow upstream of the absorbing vehicle to consist of only HDVs in this paper. Inserting CAVs to this upstream flow is expected to restrict secondary jams and is worthy of further studies.

Although we have used a single absorbing vehicle in this paper, several studies used active maneuvers of more vehicles for removing or mitigating traffic jams~\citep{Han2017,He2017}. Analysis of string stabilities in JAD scenarios with multiple absorbing vehicles will contribute the further development of JAD, and warrants future work.
%%%%%%%%%%%%%%%%%%%%%%%%%%%%%%%%%%%%%%%%%%%%%%%%%%%%%%%%%%%%%%%%%%%%%
\section*{Acknowledgements}
%%%%%%%%%%%%%%%%%%%%%%%%%%%%%%%%%%%%%%%%%%%%%%%%%%%%%%%%%%%%%%%%%%%%%
We thank Takahiro Ezaki for critical reading of the manuscript.
This work was supported by JSPS KAKENHI Grant Number JP17K14232.



%%%%%%%%%%%%%%%%%%%%%%%%%%%%%%%%%%%%%%%%%%%%%%%%%%%%%%%%%%%%%%%%%%%%%%%%%%%%%%%
% figures and tables
%%%%%%%%%%%%%%%%%%%%%%%%%%%%%%%%%%%%%%%%%%%%%%%%%%%%%%%%%%%%%%%%%%%%%%%%%%%%%%%
\clearpage

\clearpage

\clearpage
\begin{table}[t]
\caption{\label{tab:param}
Parameters of the IDM~\citep{Treiber2013}.
}
\begin{center}
\begin{tabular}{lr}
\hline
$a$ & $1\,{\rm m}/{\rm s}^2$ \\
$b$ & $1.5\,{\rm m}/{\rm s}^2$ \\
$d$ & $5\,{\rm m}$ \\
$s_0$ & $2\,{\rm m}$ \\
$v_0$ & $33.33\,{\rm m}/{\rm s}$ \\
$T$ & $1\,{\rm s}$ \\
$\delta$ & 4 \\
\hline
\end{tabular}
\end{center}
\end{table}

\clearpage

\clearpage

\clearpage

\clearpage

\clearpage

\clearpage

\clearpage

\clearpage

\clearpage

\clearpage

\clearpage
%%%%%%%%%%%%%%%%%%%%%%%%%%%%%%%%%%%%%%%%%%%%%%%%%%%%%%%%%%%%%%%%%%%%%%%%%%%%%%%
% references
%%%%%%%%%%%%%%%%%%%%%%%%%%%%%%%%%%%%%%%%%%%%%%%%%%%%%%%%%%%%%%%%%%%%%%%%%%%%%%%

% case 1: using bibtex
%%%%%%%%%%%%%%%%%%%%%%%%%%%%%%%%%%%%%%%%%%%%%
% \bibliographystyle{elsarticle-harv}
% \bibliographystyle{elsarticle-num}
% \bibliography{Jam_Removability_bibtex}
%%%%%%%%%%%%%%%%%%%%%%%%%%%%%%%%%%%%%%%%%%%%%
% case 2: using bbl
%%%%%%%%%%%%%%%%%%%%%%%%%%%%%%%%%%%%%%%%%%%%%

%%%%%%%%%%%%%%%%%%%%%%%%%%%%%%%%%%%%%%%%%%%%%

\end{document}